\pdfoutput=1
\documentclass{aa}
\usepackage{txfonts}
\usepackage{graphicx}
\usepackage{slashbox}
\usepackage{natbib}

\newcommand{\dScu}{{$\delta$~Scuti }}
\newcommand{\gDor}{{$\gamma$~Doradus }}

\bibpunct{(}{)}{;}{a}{}{,}

\begin{document}

\title{Evidence of chaotic modes in the analysis of four \dScu stars}

\author{S. Barcel\'o Forteza\inst{\ref{ins1},\ref{ins1.1}} \and T. Roca Cort\'es\inst{\ref{ins1},\ref{ins1.1}} \and A. Garc\'ia Hern\'andez\inst{\ref{ins4}} \and R.A. Garc\'ia\inst{\ref{ins3}}}

\institute{Instituto de Astrof\'isica de Canarias, 38200 La Laguna, Tenerife, Spain \label{ins1}
\and
Departamento de Astrof\'isica, Universidad de La Laguna, 38206 La Laguna, Tenerife, Spain \label{ins1.1}
\and
Instituto de Astrof\'isica e Ci\^encias do Espaço, Universidade do Porto, CAUP, Rua das Estrelas, PT4150-762 Porto, Portugal\label{ins4}
\and
Laboratoire AIM, CEA/DRF – CNRS – Univ. Paris Diderot – IRFU/SAp, Centre de Saclay, 91191 Gif-sur-Yvette Cedex, France\label{ins3}
}

\date{Received 8 April 2016; Accepted 24 February 2017}

\abstract{Since CoRoT observations unveiled the very low amplitude modes that form a flat plateau in the power spectrum structure of \dScu stars, the nature of this phenomenon, including the possibility of spurious signals due to the light curve analysis, has been a matter of long-standing scientific debate.}{We contribute to this debate by finding the structural parameters of a sample of four \dScu stars, CID~546, CID~3619, CID~8669, and KIC~5892969, and looking for a possible relation between these stars' structural parameters and their power spectrum structure.}{For the purposes of characterization, we developed a method of studying and analysing the power spectrum with high precision and have applied it to both CoRoT and Kepler light curves.}{We obtain the best estimates to date of these stars' structural parameters. Moreover, we observe that the power spectrum structure depends on the inclination, oblateness, and convective efficiency of each star.}{Our results suggest that the power spectrum structure is real and is possibly formed by 2-period island modes and chaotic modes.}


\maketitle

\section{Introduction}
\label{s:intro}

The launch of space telescopes such as MOST, CoRoT, \& Kepler satellites \citep{Walker2003,Baglin2006,Borucki2010} announced the beginning of the precise study of the stellar oscillations in stars other than the Sun. Since then, the high quality of the light curves has allowed the precise characterization of the mode parameters of different kinds of stars, and the study of their variation with time and their connection with the stellar structure.\\

Although the power-spectral structure of the stars with solar-type oscillations is well known, this is not the case for \dScu stars. The power spectrum of these stars shows a complex structure with dominant peaks of moderate amplitudes and many hundreds of lower amplitude peaks that form a flat plateau \cite[e.g.][]{Poretti2009}, the so-called "grass". After observation of the "grass", a long-standing debate about its origin started, including the possibility of it arising from spurious signals produced during the analysis of the data \citep{Balona2014a}.\\

A huge theoretical effort has been made to find a possible physical phenomenon behind this power-spectral structure. Some of these arguments are:\\
1. Less effective disc-disc averaging of the flux owing to the geometry of the \dScu star. Therefore, it is possible to find modes with higher degrees than in the spherical symmetric case $l > 4$ \citep{Balona1999}. Although \citet{Balona2011} find that most of \dScu stars do not seem to have enough density of peaks to support this possibility, several stars seem to show modes with high degrees, up to $l = 20$ \citep{Kennelly1998,Poretti2009}.\\
2. A granulation background signal due to the effect of a thin outer convective layer \citep{Kallinger2010}. This effect is found to be more important in cool \dScu stars \citep{Balona2011a}.\\
3. Variations with time that produce sidelobes of the main peak of the spectra. \citet{Balona2011} find that around $\sim$45\% of the spectra of Kepler \dScu stars have one-sided sidelobe. They discard effects such as binarity because these yield amplitude-symmetric equal-spaced multiplets \citep{ShK2012}. However, there are other causes that produce variations with non-symmetric amplitude multiplets, such as resonant mode coupling \citep[RMC; see][and references therein]{BarceloForteza2015}.\\
4. A magnetic field in a rotating star splits each peak of the rotational multiplet into $(2l+1)$, meaning that one mode is split into $(2l+1)^{2}$ peaks \citep{Goode1992}. Magnetic fields have been detected in the surface of $\sim$7\% of main sequence and pre-main sequence intermediate-mass and massive stars \citep{Mathis2015}. However, \dScu stars with measurable magnetic fields are not common because only one \dScu star shows a magnetic field \citep{Neiner2015} and another is suggested to be magnetic from its chemical abundance \citep{Escorza2016}.\\
5. The oblateness of the star produced by high rotation rates is the cause of the appearance of a significant number of chaotic modes \citep{Lignieres2009}.\\

The determination of the fundamental structural parameters of these stars such as mass, inclination, rotation rate, and convective efficiency can help us to unveil which of these mechanisms are responsible for this kind of power spectral structure. Four interesting \dScu stars observed by CoRoT and Kepler satellites that have been characterized in this paper are CID~546\footnotemark[1]\footnotetext[1]{Stars observed by CoRoT have a CoRoT ID of ten numbers. We abbreviate them by not taking into account those left-sided zeros. For example, CoRoT ID~0000000546 is called CID~546.}, CID~3619\footnotemark[1], CID~8669\footnotemark[1], and KIC~5892969. Their differences in the power spectra help us in our aim. In Sect.~\ref{s:dScu} we describe the main characteristics of these kind of stars. The way in which the their oscillations are analysed to obtain the parameters of the modes is presented in Sect.~\ref{s:dSBF}. Results for each target star are commented in Sect.~\ref{s:4dScu}. In Sect.~\ref{s:regular} we estimate their structural parameters. The power-spectral structure is deeply studied and discussed in Sect.~\ref{s:grass}. In the last section we present our conclusions.

\section{\dScu type stars}
\label{s:dScu}

\dScu stars are classical pulsators with oscillation frequencies between $\sim$60 and $\sim$900 $\mu$Hz \citep[e.g.,][]{Zwintz2013}. These stars are located on or slightly off the main sequence, with spectral types between A2 and F5 \citep{Breger2000}. They are intermediate-mass stars that show fast rotation rates as it is common in stars within their mass domain or of higher mass \citep{Royer2007}. In fact, one of the reasons that \dScu stars can be separated from RR~Lyrae stars is their higher rotational velocity, $v \mathrm{sin}i > 10$ km/s \citep[see][]{Peterson1996}. Other typical characteristics of \dScu stars are detailed in Table \ref{t:dScuchar}.\\

\begin{table}
\caption{Typical values of the stellar characteristics of \dScu stars by \cite{Breger2000}, \cite{Aerts2010}\tablefootmark{a}, and \cite{Uytterhoeven2011}\tablefootmark{b}}
\label{t:dScuchar}
\centering
\begin{tabular}{c c c c}
\hline \hline  
Characteristic                                              & From & To \\ \hline
Spectral-type                                               & F5   & A2 \\ 
Luminosity class                                            & III  & V \\ 
$M$ ($M_{\odot}$)                                           & 1.5  & 2.5 \\ 
$T_{\mathrm{eff}}$ (K) \tablefootmark{b}                    & 6300 & 8600 \\ 
$\mathrm{log} ~\textit{g}$ (cgs)\tablefootmark{b}           & 3.2  & 4.3 \\
$v \mathrm{sin}i$ (km/s)                                    & 10   & 250 \\
$\nu$ ($\mu$Hz)                                             & 60   & 930\tablefootmark{a} \\
A (mag)                                                     &      & 0.3 \\
\hline 
\end{tabular}
\end{table}  

\subsection{Hybrid stars}
\label{ss:subgroups}

Several subgroups can be distinguished from the main class of \dScu stars pulsating with nonradial p-modes, such as High Amplitude \dScu stars (HADS), SX~Phe variables, or $\delta$~Scu/$\gamma$~Dor hybrid stars \citep{Breger2000}.  This last group comes from the observation of g-modes in \dScu type stars with frequencies typical of \gDor stars, meaning $\nu \sim \left[6-60 \right] \mu$Hz. \citet{Uytterhoeven2011} point out that a star can be classified as hybrid when all of the three following conditions are accomplished:\\
1) Typical frequencies of both kinds of stars are detected.\\
2) The amplitudes of both domains are comparable, within a factor $\lesssim 5$.\\
3) There are two independent frequencies in both domains with amplitudes higher than 100 parts per million (ppm)\\
If the star is hybrid it would be a $\delta$~Scu/$\gamma$~Dor or a $\gamma$~Dor/$\delta$~Scu star depending on which part is the dominant one \citep{Grigahcene2010}. In this way, it is found that a great amount of \dScu and \gDor stars are hybrids, $\sim$36\%. Other studies suggest that all \dScu stars are hybrids \citep{Balona2014}.\\

However, other magnitudes that help us to differentiate \dScu from \gDor are the convective efficiency ($\Gamma$),
\begin{equation}
\Gamma \sim \left(T_{eff}^{3} \mathrm{log} ~\textit{g} \right)^{-\frac{2}{3}}\, ,
\label{e:conveff}
\end{equation}
where $g$ is the surface gravity, and the kinetic energy of the waves ($E_{kin}$),
\begin{equation}
E_{kin} \sim \left(A_{0} \nu_{0} \right)^{2}\, ,
\label{e:kine}
\end{equation}
where $\nu_{0}$ and $A_{0}$ are the frequency and amplitude of the mode with maximum power, respectively. These magnitudes have a dominant value of $\mathrm{log}~\Gamma<-8.1$ and $\mathrm{log}~E_{kin}>10.1$ for \dScu stars when the amplitude is measured in ppm and the frequency in $\mu$Hz \citep[see][]{Uytterhoeven2011}. Both quantities are related to the convective zone of the star which is more efficient in \gDor stars. We use all of these tools to find out whether some of our selected stars are hybrid stars and if they present some other differences in their power-spectral structure.\\

\subsection{Rotational effect}
\label{ss:rot}

Taking into account the effect of rotation on the perturbation analysis of a spherically symmetric star, the modes split into multiplets. For low rotation rates ($\Omega$), these multiplets present $(2l+1)$ symmetric peaks, split approximately a multiple of the rotational splitting ($s$). For higher rotation rates, second-order effects have to be taken into account and the symmetry is broken \citep{Saio1981,Dziembowski1992}. Besides the appearance of the asymmetry, the multiplet is also globally shifted. To go even further, a third-order correction has already been studied by \citet{Soufi1998}.\\

The different contributions can be estimated thanks to two dimensionless magnitudes \citep{Goupil2000} as
\begin{equation}
\epsilon^{2} = \frac{\Omega^{2} R^{3}}{G M} \, , \qquad \mu = \frac{\Omega}{2 \pi \nu} \, ,
\label{e:eps2}
\end{equation}
where $\epsilon$ scales the effect of centrifugal force with gravity and $\mu$ scales the effect of the rotational rate to oscillation frequencies. All of these effects are higher for lower frequency g-modes than for higher frequency p-modes. However, they can produce observable shifts up to 1 $\mu$Hz.\\

Rotation also produces a deformation of the star \citep{Cassinelli1987}. Under the assumption that the rotation is uniform and the surface of the star is approximately a Roche surface \citep{PerezHernandez1999}, an averaged effective gravity ($g_{eff}$) can be defined as
\begin{equation}
g_{eff} = g - \frac{2}{3} R \Omega^{2}\, ,
\label{e:geff}
\end{equation}
where $R$ is the radius of the star with spherical symmetry. With these assumptions it is also possible to obtain the polar radius $R_{p}$,
\begin{equation}
R_{p} = \frac{R}{1+\frac{\epsilon^{2}}{3}}\, .
\label{e:Rp}
\end{equation}
Assuming that the volume is constant compared with a spherically symmetric star, the oblateness $O$ of the star is also defined as
\begin{equation}
O = 1-\left( 1+ \frac{\epsilon^{2}}{3} \right)^{-\frac{3}{2}} \; .
\label{e:oblateness}
\end{equation}
The oblateness of a star increases with higher $\epsilon$. However, rotation has a maximum limit at which the centrifugal force will destroy the star. This limit is the so-called break-up frequency ($\Omega_{K}$):
\begin{equation}
\Omega_{K} = \left( \frac{8 G M}{27 R_{p}^{3}} \right)^{\frac{1}{2}}
\label{e:Omegak}
\end{equation}
Therefore, for a stable \dScu star, $\Omega/\Omega_{K} \leq 1 $ has to be accomplished.\\

Another known effect of the rotation is gravity darkening \citep{vonZeipel1924}, meaning an increase of the temperature from the equator to the poles. It follows a potential law as
\begin{equation}
\delta T_{eff} = \frac{T_{eff,p}-T_{eff,e}}{T_{eff,p}} \approx 1-\left( 1- \epsilon^{2} \right)^{\frac{\beta}{4}} \; ,
\label{e:dteff}
\end{equation}
where the value of $\beta$ depends of the evolutionary stage of the star \citep{Claret1998}. Taking into account this effect, the measurements of $T_{eff}$ and $\mathrm{log} ~\textit{g}$ will vary depending on the inclination angle of the star ($i$). Therefore, these variations have to be carefully treated to obtain the proper main characteristics of the star.

\section{Analysis of CoRoT \& Kepler \dScu light curves}
\label{s:dSBF}

The data we use were obtained by the CoRoT and \textit{Kepler} satellites. The CoRoT satellite (Convection, Rotation, and planetary Transits) was developed and operated by the French space agency CNES with international contributions of ESA, Austria, Belgium, Brazil, Germany and Spain. The objective of the mission was to search for exoplanets and to perform asteroseismic studies. Two different channels were designed: the exo-channel data are sampled every 512 s and their photometric precision is between 40 and 90 ppm \citep{Auvergne2009}, whereas the seismo-channel has a much shorter cadence of 32 s and a substantially higher photometric precision of between 0.6 and 4 ppm \citep[see][]{Auvergne2009}. As described in \cite{Boisnard2006}, two kinds of campaigns were planned of different duration: Long Runs (LR) of approximately 150 d, and Short Runs (SR) with durations around 30 d. These runs were carried out in two different fields: one close to the galactic anticentre direction(a) and the other close to the galactic centre direction (c). The light curves of each star can then be classified with the nomenclature <duration><direction><number>. For example, LRa01 is the first Long Run to point close to the anticentre of the galaxy.\\

The \textit{Kepler} mission was designed by NASA to survey a single region of our own galaxy to detect and characterize Earth-sized planets close to their habitable zones using the transit method \citep{Thompson2012}. The maximum duration of their light curves are up to four years. Because the original field is no longer observable, the mission was renamed as \textit{K2} \citep{Howell2014}. There are two different cadences available depending on the star: Long Cadence (LC) of $\sim$29.5 min or the Short Cadence (SC) of $\sim$1 min. Moreover, the data are downloaded in three-months bases called quarters \citep[Q<number>,][]{Haas2010}.\\

The asteroseismic analysis was performed by a set of different programs called \dScu Basics Finder ($\delta$SBF) that we built using IDL language programming. We analysed the light curves of \dScu stars with the three stage Method described in detail in \citet{BarceloForteza2015}. This iterative method allows us to interpolate the light curve using the information of the subtracted peaks, minimizing the effect of gaps, considerably improving the background noise, and avoiding spurious effects \citep{Garcia2014}. Thus, we get very accurate results in terms of frequencies, amplitudes, and phases. In addition, we take into account the energy of the signal for each peak $i$:
\begin{equation}
E_{i} =\frac{RMS_{i}-RMS_{i+1}}{RMS_{0}} \, ,
\label{e:esignal}
\end{equation}
where $RMS_{i}$ is the root mean square of the residual signal after the subtraction of the highest $i$ peaks, and for which i=0 is the original signal.\\

The next step in our strategy was to characterize each \dScu star, finding its structural parameters such as mass ($M$) or oblateness ($O$). This is possible thanks to the regularities present in the power-spectral structure of these kind of stars, such as the large separation ($\Delta\nu$) and the rotational splitting ($s$).\\ 

The last step consists in studying each star's power-spectral structure in order to find any relation between their characteristic parameters, such as density of peaks ($\mathbf{n}_{mean}$), and their structural parameters. These two steps will be further discussed in the sections that follow.\\

\section{Characterization of four \dScu stars}
\label{s:4dScu}

We tested our method with an already known \dScu star: CID~546. We also analysed the light curves of CID~3619, CID~8669, and KIC~5892969 stars. Figures~\ref{f:S_4dScu_546} to~\ref{f:S_4dScu_5893969} show the four power spectra and their frequency contents. The highest amplitude peaks can be found at Appendix~\ref{ap:PoM}.\\

\subsection{CID~546}
\label{ss:test}

\begin{figure}[!t]
\centering
\includegraphics[width=0.495\textwidth]{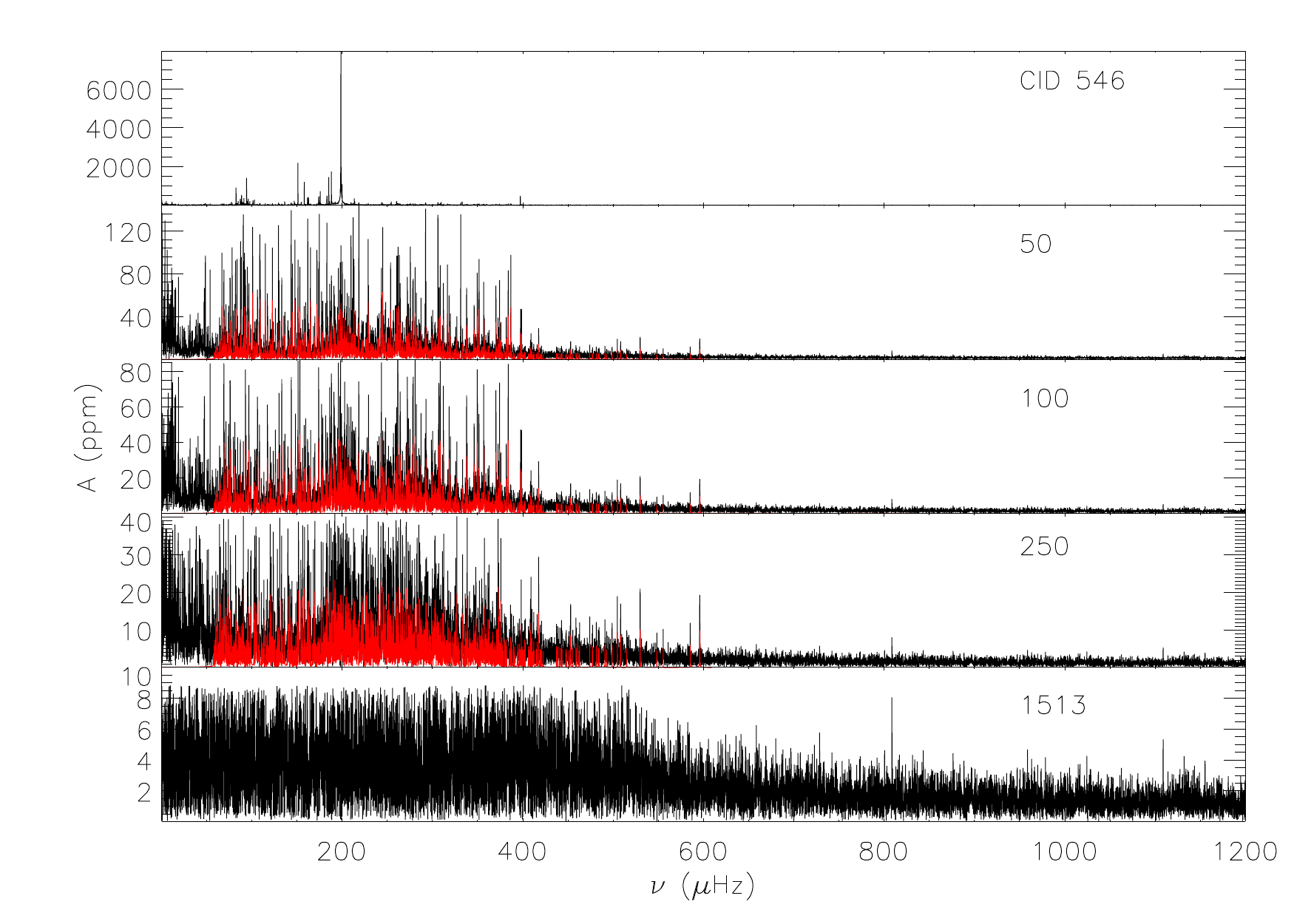} \\
\protect\caption[]{From top to bottom: Power-spectral structure of the original light curve for CID~546, and also those after extracting the indicated number of peaks. The contribution of the peaks considered as "grass" is represented in red (see Sect.~\ref{s:grass}).}
\label{f:S_4dScu_546}
\end{figure}

\begin{figure}[!t]
\centering
\includegraphics[width=0.495\textwidth]{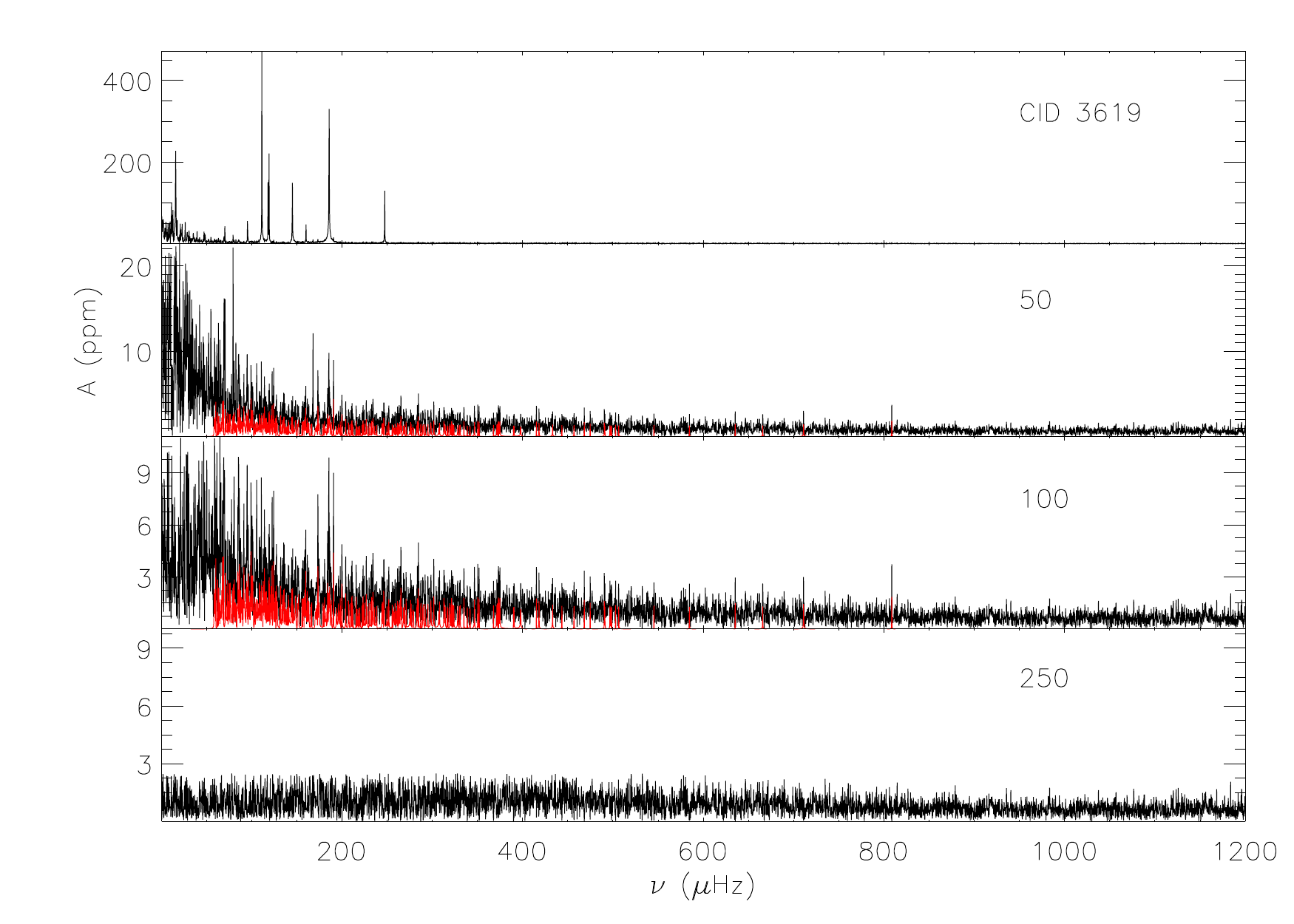}
\protect\caption[]{Same as Fig.~\ref{f:S_4dScu_546} for CID~3619.}
\label{f:S_4dScu_3619}
\end{figure}

\begin{figure}[!t]
\centering
\includegraphics[width=0.495\textwidth]{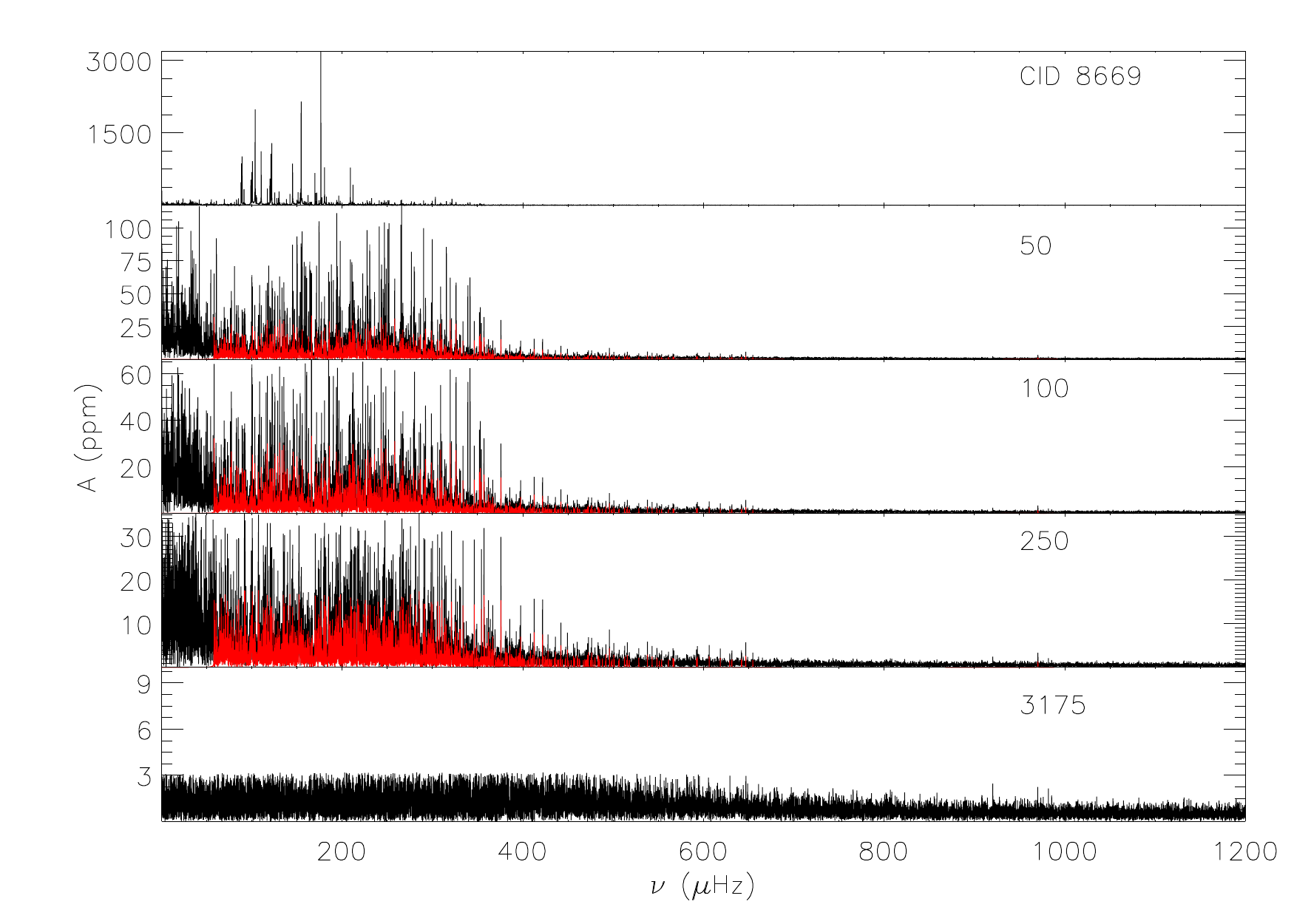}
\protect\caption[]{Same as Fig.~\ref{f:S_4dScu_546} for CID~8669.}
\label{f:S_4dScu_8669}
\end{figure}

\begin{figure}[!t]
\centering
\includegraphics[width=0.495\textwidth]{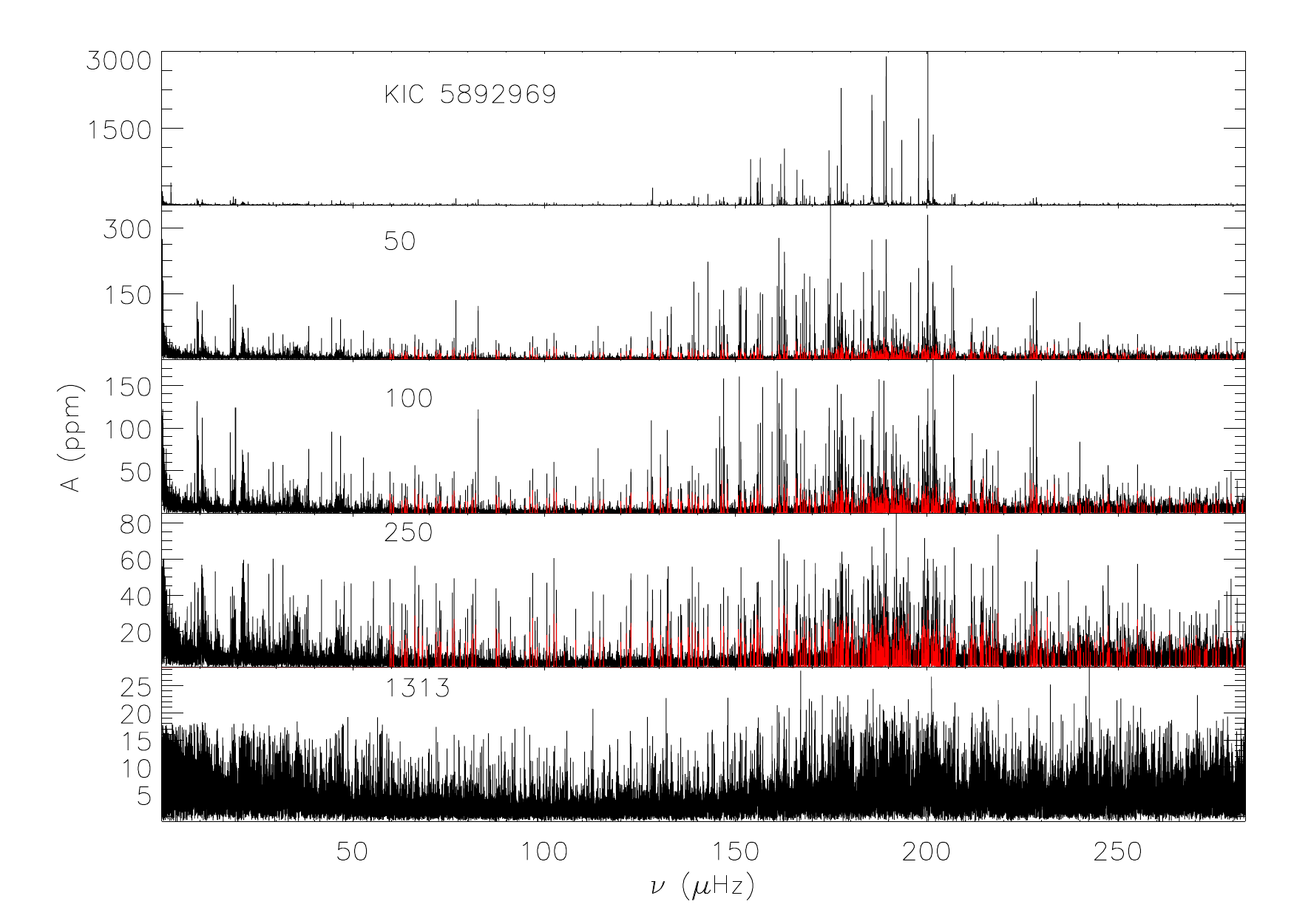}
\protect\caption[]{Same as Fig.~\ref{f:S_4dScu_546} for KIC~5892969.}
\label{f:S_4dScu_5893969}
\end{figure}

The \dScu star CID~546 (HD~50870) is a F0IV star with $M_{v} \sim$1.67 at an approximate distance of 277 pc \citep[$V\sim8.88$;][]{McCuskey1956}, observed by CoRoT close to the anticentre direction of the Galaxy. It has been observed during 114.4 d between 2008 November 13 and 2009 March 8 (LRa02). Its known parameters are listed in the first rows of Table~\ref{t:4dScu}. Recently, \citet{Mantegazza2012} discovered that CID~546 is a long-period spectroscopic binary star with a cooler companion. Nevertheless, additional observations that cover more of the orbital period are necessary to confirm their results, which they consider to be preliminary. No evidence of binarity is found in the photometric analysis.\\

When our analysis is applied to the power spectrum of this star, we obtain 1513 peaks higher than 10 ppm with a signal-to-noise ratio (SNR) greater than or equal to four. These peaks carry 99.77\% of the full signal. Seventeen peaks with amplitudes higher than 400 ppm carry 97.37 \% of the signal and appear in three different frequency ranges (see Table~\ref{t:peaks546} and top panel of Fig.~\ref{f:S_4dScu_546}): the main regime from 150 to 200 $\mu$Hz; the second, which is half the value of the previous one, from 80 to 100 $\mu$Hz, approximately; and the third regime close to 400 $\mu$Hz. The analysis also reveals 1446 peaks with amplitudes lower than 130 ppm that only carry 0.95\% of the energy of the signal. The flat plateau is clearly visible and could be differentiated from noise after extracting hundreds of peaks (see Fig.~\ref{f:S_4dScu_546}).\\

Comparing the results with those obtained by \citet{Mantegazza2012}, our method finds all peaks with amplitudes higher than 50 ppm and frequencies higher than 1 $\mu$Hz, with mean relative differences between both results of $5\times 10^{-3}$\% in frequency and 10\% in amplitude. The differences between frequencies obtained for both methods are $\sim$0.01 $\mu$Hz, one order of magnitude lower than the frequency resolution. Therefore, it can be concluded that the method produces accurate results and allows us to study real light-curves of \dScu stars from both the CoRoT and Kepler satellites.\\

\subsection{CID~3619}
\label{ss:chocobo}

The F0V star CID~3619 (HD~48784) has $M_{v} \sim$1.87 at a distance of approximately 91 pc \citep[$V\sim6.65$;][]{Charpinet2006} and was observed by CoRoT close to the anticentre direction of the Galaxy. The satellite followed it for 25.3 d during 2008 March 5-31 (SRa01) and also for 40.7 d during 2011 November 29 - 2012 January 9 (SRa05).\\

The analysis of the power spectrum of the SRa01 light curve detects 163 peaks down to 5 ppm with a SNR greater than or equal to four. These peaks carry 96.45\% of the full signal. The same analysis was performed for the SRa05 light curve obtaining 508 peaks down to 2.5 ppm with the same lower limit for the SNR and carrying 96.8 \% of the full signal. We found 37 and 42 peaks with energies higher than 0.1\%, respectively. Of all these peaks, only twenty are detected in both runs (see Table~\ref{t:peaks3619}), which show slight differences in frequency, up to 0.2 $\mu$Hz, but higher changes in amplitude, up to the 73\%, and phase, up to 1.5$\pi$.\\

The power-spectral structure of this star shows a frequency range that includes the typical regimes of both \gDor and \dScu stars. Moreover, the ratio between mean amplitudes of both regimes is around $A_{\delta Scu}/ A_{\gamma Dor} \sim$4, which means that CID~3619 is a hybrid $\delta$~Scu/$\gamma$~Dor star candidate. The power spectrum of CID~3619 does not show the flat plateau present for CID~546 (see Figures~\ref{f:S_4dScu_546} and~\ref{f:S_4dScu_3619}).\\

\subsection{CID~8669}
\label{ss:cactuar}

The A5 \dScu star CID~8669 (HD~181555) is of absolute magnitude $M_{v} \sim$2.19 at a distance of approximately 116 pc \citep[$V\sim7.52$;][]{Charpinet2006} and was observed by CoRoT close to the direction of the centre of the Galaxy. It was observed for 156.6 d between the 2007 May 11 and 2007 October 15 (LRc01).\\

The power spectrum of this star shows 3175 peaks higher than 3 ppm with a SNR greater o equal to four. These peaks carry 99.83\% of the full signal. Thirty-one peaks with amplitudes higher than 200 ppm carry 95.71 \% of the energy of the signal (see Table~\ref{t:peaks8669}). The analysis also finds 3054 peaks with amplitudes lower than 70 ppm that only carry a 1.75\% of the energy of the signal. The flat plateau is also visible and it has a higher density of peaks than CID~546 (see Figures~\ref{f:S_4dScu_546} and~\ref{f:S_4dScu_8669}).\\

\subsection{KIC~5892969}
\label{ss:tomberi}

The stellar characteristics of the faint \dScu star KIC~5892969, $Kp\sim$12.445, have been studied spectroscopically by \citet{Huber2014}. This star has been observed by the Kepler satellite during 1470 d, from Q0 to Q17, in LC and its oscillation modes have been studied in \citet{BarceloForteza2015}. Since the Nyquist frequency of KIC~58929's power spectrum is lower than the typical frequency range of \dScu stars, we cannot ascertain a priori whether there is a flat plateau (see Fig.~\ref{f:S_4dScu_5893969}).\\

\section{Searching for possible spectral regularities}
\label{s:regular}

\begin{figure*}[!t]
\centering
\includegraphics[width=0.495\textwidth]{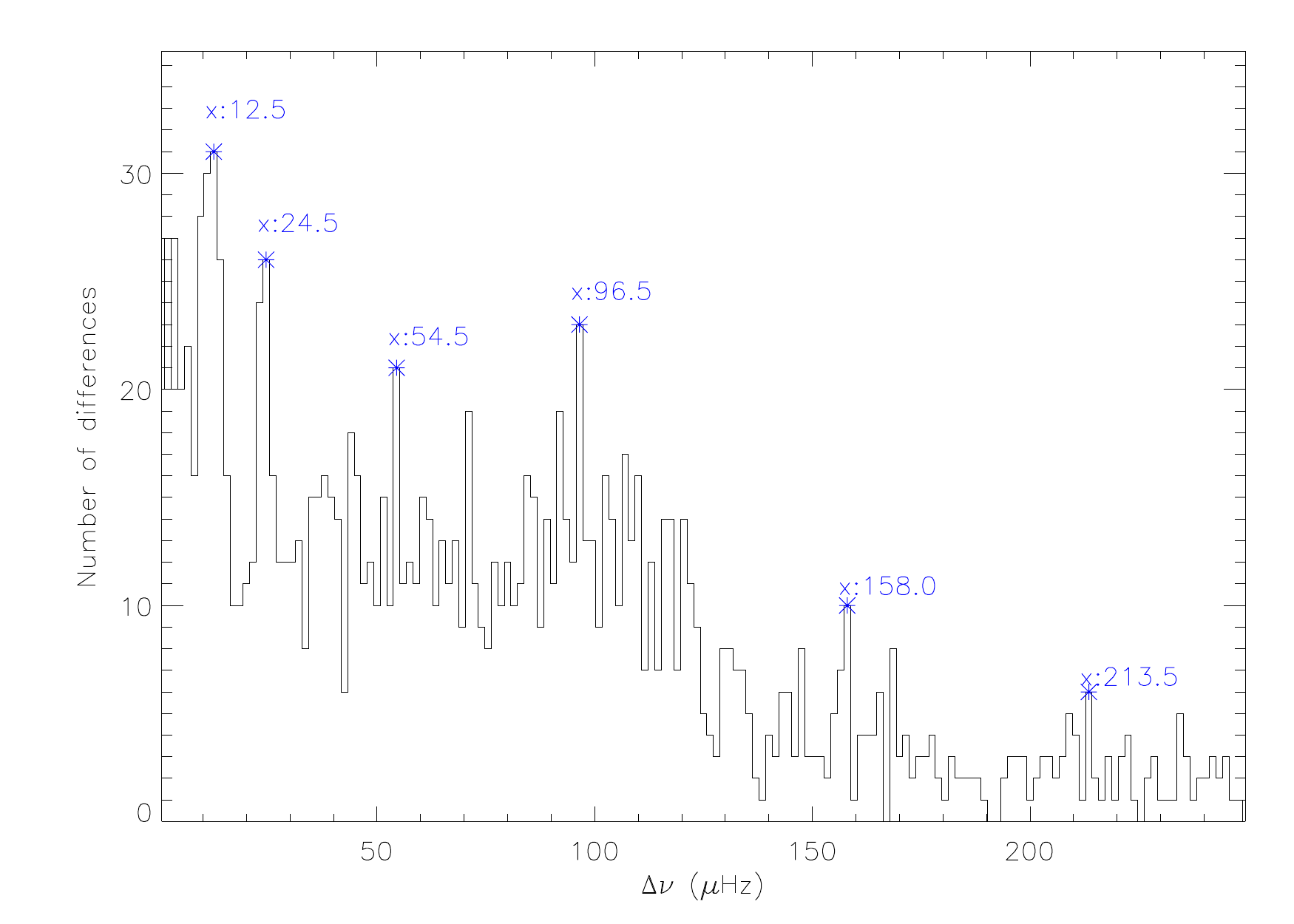}
\includegraphics[width=0.495\textwidth]{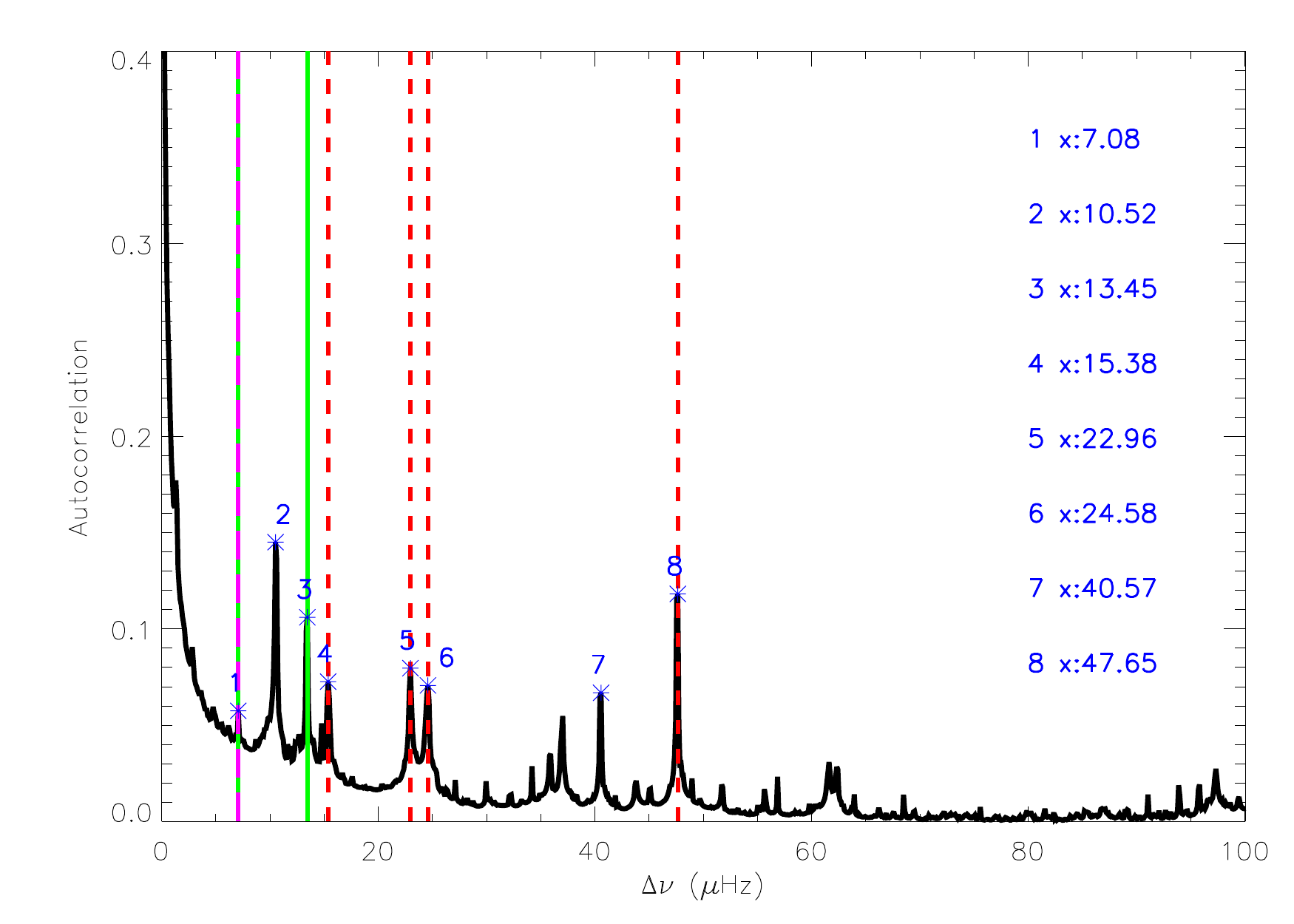}
\includegraphics[width=0.495\textwidth]{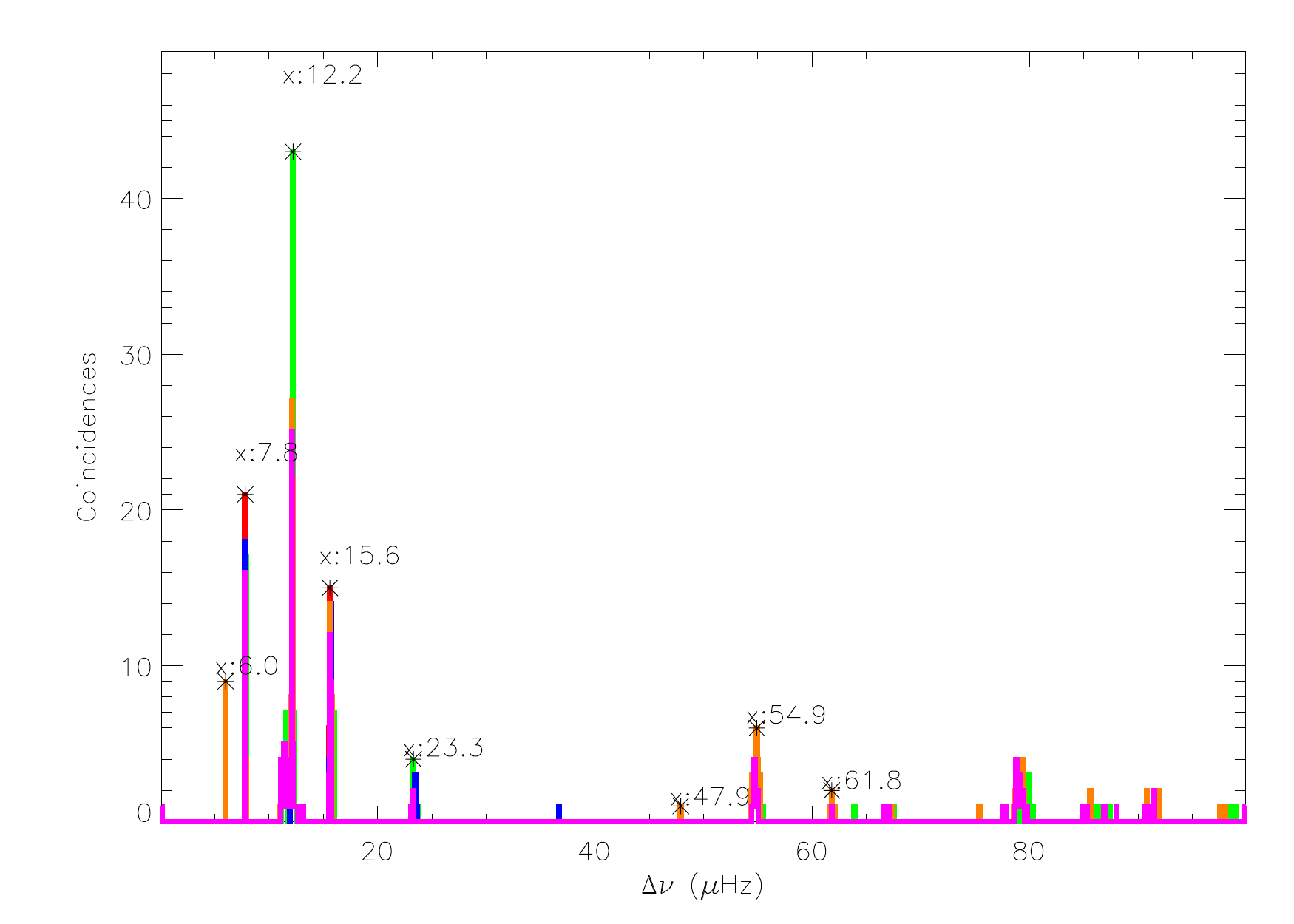}
\includegraphics[width=0.495\textwidth]{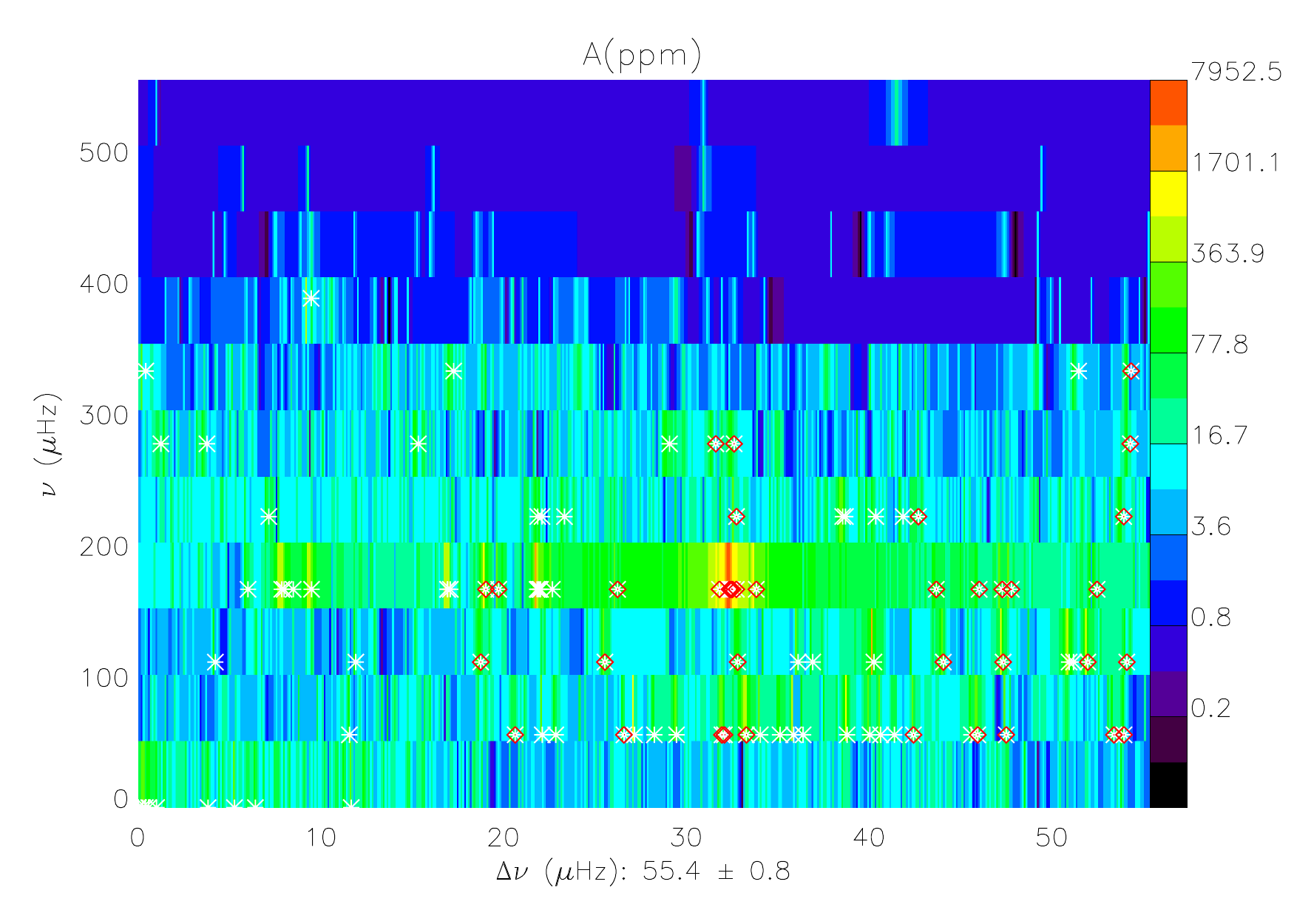}
\protect\caption[]{\textit{Top left panel}: Histogram of differences of CID~546 modes. \textit{Top right panel}: Autocorrelation function of CID~546 power spectra. Red dashed lines point to the large separation or one of its split peaks and its submultiples. Purple long-dashed line point to the splitting calculated by the difference between the supposed large separation and the nearest peak. This difference coincides with the rotational signature found and its double both marked with green solid lines. \textit{Bottom left panel}: Histogram of coincidences for possible regularities in CID~546 power spectrum (red). Other colours point to the coincidences achieved with an artificial spectrum built taking into account different number of highest amplitude peaks ($I$): 50 (green), 100 (blue), 250 (orange), and all detected peaks (purple). \textit{Bottom right panel}: Echelle diagram for CID~546 power spectra using a periodicity of 55.4 $\mu$Hz. White asterisks correspond to the highest amplitude modes and red diamonds represent those consecutive modes that are aligned. Notice the logarithmic scale for the amplitude axis.}
\label{f:S_C546}
\end{figure*}

As \citet{Suarez2014} stress, the mode organization for \dScu stars includes regularities as the large separation with a negligible variation from the non-rotating case. Using a dense sample of representative models, they obtain the following scaling relation:
\begin{equation}
\frac{\Delta\nu}{\Delta\nu_{\odot}} = 0.776\left( \frac{\rho}{\rho_{\odot}}\right)^{0.46}\, ,
\label{e:lsepsuarez}
\end{equation} 
where $\rho$ is the mean density of the star. This relation is somewhat similar to that found to solar-type oscillators \citep{Kjeldsen1995}:
\begin{equation}
\frac{\Delta\nu}{\Delta\nu_{\odot}} = \left( \frac{\rho}{\rho_{\odot}}\right) ^{\frac{1}{2}}\, .
\label{e:lsep}
\end{equation} 
In fact, they point out that the minimum error of these analyses is around 11 to 21\% and is due to the stellar deformation. This is in agreement with several previous theoretical studies claiming that regularities in the p-modes of \dScu stars are related to the spherical large separation \citep[e.g.][]{Pasek2012}.\\

On the observational side, many \dScu stars show frequency spacings \citep[e.g.:][]{GarciaHernandez2009,Zwintz2011a}. Some of these spacings have been interpreted as a combination of frequencies \citep{Breger2011} or the signal of the rotational splitting \citep{Zwintz2011}. However, several \dScu stars known to be eclipsing binaries have been analysed \citep[e.g.:][]{daSilva2014}. As binary stars, it is possible to calculate their stellar characteristics such as mass or radius. With all these data, \citet{GarciaHernandez2015} find a similar relation to the previous one proving that the above scaling relation is independent of the rotation rate.\\

Therefore, using previously known parameters, the $\Delta\nu - \rho$ relation, and considering
\begin{equation}
\frac{v \sin i}{2 \pi R} \leq s \leq \frac{\Omega_{K}}{2 \pi} \; ,
\label{e:slim}
\end{equation}
it is possible to delimit the value of these two regularities. Once we have the mean density of the star and its rotation, the mass and the radius can be estimated using the Stephan-Boltzmann law and the surface gravity acceleration (Eq.~\ref{e:geff}). Moreover, we can also obtain a mass estimate with the mass-luminosity relation \citep{Ibanovglu2006}.\\

We used the following four methods to look for regularities.\\

\subsection{Histogram of differences}
\label{ss:HD}

\begin{table*}
\caption{Obtained values of splitting and large separation for each method and star.}
\label{t:sdnumethod}
\centering
\begin{tabular}{c | c c c c | c c c c}
            & \multicolumn{4}{c}{Splitting} &  \multicolumn{4}{c}{Large Separation}\\
\hline
Star & HoD\tablefootmark{1} & AC\tablefootmark{2}& TRUFAS\tablefootmark{3}& ED\tablefootmark{4}& HoD\tablefootmark{1} & AC\tablefootmark{2} & TRUFAS\tablefootmark{3} & ED\tablefootmark{4} \\
\hline \hline  
KIC~5892969 &1.1 $\pm$ 0.2&1.3 $\pm$ 0.1&1.2 $\pm$ 0.3 &  1.2 $\pm$ 0.1 & 22.2 $\pm$ 0.5 & 22.6 $\pm$ 0.9 & 22.1 $\pm$ 0.2 &22.2 $\pm$ 0.4\\
CID~546     &  6.4 $\pm$ 0.6 &  7.5 $\pm$ 0.6 &  7.0 $\pm$ 1.1 &  7.1 $\pm$ 0.2 & 53 $\pm$ 3 & 54 $\pm$ 1 & 55 $\pm$ 1 & 55.4 $\pm$ 0.8 \\ 
CID~3619    &  8.6 $\pm$ 0.4 &  8.1 $\pm$ 0.3 &  7.6 $\pm$ 0.7 &  8.1 $\pm$ 0.5 & 41 $\pm$ 3 & 40 $\pm$ 2 & 40 $\pm$ 2 & 40.3 $\pm$ 0.6 \\ 
CID~8669    & 16.3 $\pm$ 0.4 &  18 $\pm$ 1    & 16.7 $\pm$ 0.1 & 16.9 $\pm$ 0.7 & 55 $\pm$ 1 & 54 $\pm$ 1 & 55 $\pm$ 2 & 55.0 $\pm$ 0.6 \\ 
\hline 
\end{tabular}
\tablefoot{Each column points to the results of each method: \tablefoottext{1}{The histogram of diferences (see Section~\ref{ss:HD})},\tablefoottext{2}{the autocorrelation function (see Section~\ref{ss:AC})},\tablefoottext{3}{the spectrum of the subspectrum analysis (see Section~\ref{ss:deepground})}, and \tablefoottext{4}{the echelle diagram (see Section~\ref{ss:ED}).}}
\end{table*}

\citet{Breger2009} use a histogram of differences between all the detected modes and find that the radial modes ($l=0$) are not the only kind of mode that allow us to find the large separation. \citet{GarciaHernandez2009} stress that high amplitude modes carry this signature, and including the lower amplitude modes powers other periodicities. These other regularities make it very difficult to determine the large separation. Therefore, we used the histogram of differences between pairs of frequencies within the typical \dScu star range of frequency oscillations and only taking into account the, approximately, 50 highest amplitude peaks (see top left panel in Fig.~\ref{f:S_C546}).\\

Rotational multiplets can be present in the set of frequencies chosen for the analysis, allowing us to also see the rotational splitting. As we mention in Sect.~\ref{ss:rot}, the higher the rotation rate, the greater the deviation from a symmetric splitting, and the more difficult it is to observe in the histogram \citep{Goupil2000}. For low rotation rates, $s \sim 1$ $\mu$Hz, the splitting is easily detectable and contributions due to twice and thrice the splitting are also detected. Moderate rotation rates, $s \sim 5$ $\mu$Hz, perturb this structure but the splitting is still dominant. For higher rotation rates, $s \sim 10$ $\mu$Hz, it is not possible to observe a dominant peak, owing to the lack of symmetry between the peaks of the rotational multiplet.\\

To obtain an accurate value of these parameters with this method, it is important that the binning of the histogram reaches a compromise between the possible variation with frequency of the periodicities and the accuracy we want to reach. We used a bin of 0.5 $\mu$Hz to look for rotational signatures and bins up to 1.5 $\mu$Hz to find the large separation. The results of this method (see Table~\ref{t:sdnumethod} and top left panel of Fig.~\ref{f:S_C546}) take into account possible multiples of the rotation, multiples or submultiples of the large separation, and their split peaks ($\Delta \nu \pm s$).\\

\subsection{Autocorrelation function}
\label{ss:AC}

The autocorrelation function compares a power spectrum with itself as a function of lag. This function has a higher value when the variations of the original spectrum increase and decrease similarly to the shifted spectrum. Then, when the lag coincides with one of the possible regularities of the spectrum, the value of the autocorrelation function increases.\\

\citet{Reese2013} test this method with artificial spectra, taking into account a different number of modes and calculating their visibilities with different inclinations and rotation rates. They conclude that it is possible to obtain regularities corresponding to the large separation and half its value, and also the rotation rate and twice its value. These peaks are reinforced when the range of observed frequencies spans a large enough interval and does not include too many modes in the artificial light-curve. This last condition is important to avoid powering other regularities that can be present in the spectrum, as also happens with the histogram of differences. Therefore, we calculated the autocorrelation function of the artificial spectrum that is built, by considering only the approximately 50 highest amplitude modes (see top right panel in Fig.~\ref{f:S_C546}).\\

Once the autocorrelation function was calculated, we looked for its highest values in a 1 to 100 $\mu$Hz lag interval. The large separation is found by looking for the peak with highest number of consecutive submultiples within its error. Then, we looked for its closest peaks to find possible rotation signatures. The last step was to try to find the rotation signature by looking for one peak within the 1 to 18 $\mu$Hz range that has consecutive multiples within its error. This method usually finds one of the split peaks of the large separation as the dominant peak ($\Delta \nu \pm s$). We correct it by adding or subtracting the rotational splitting (see Table~\ref{t:sdnumethod} and top right panel of Fig.~\ref{f:S_C546}).\\

\subsection{TRUFAS: the spectrum of a subspectrum}
\label{ss:deepground}

\begin{table*}[!t]
\caption{Previously known and calculated parameters of four CoRoT and Kepler \dScu stars.}
\label{t:4dScu}
\centering
\begin{tabular}{c c c c c c}
\hline\hline
       & Stars                   & KIC~5892969          & CID~546           & CID~3619                     & CID~8669          \\\hline
 Already   & $T_{eff}$(K)&  7560 $\pm$ 250      & 7600 $\pm$ 200    & 6990 $\pm$ 140               & 7000 $\pm$ 200    \\
 known     & $\mathrm{log}~\textit{g}_{eff}$& 3.8 $\pm$ 0.3& 3.9 $\pm$ 0.2             & 4.0       & 4.3  $\pm$ 0.2     \\
 parameters\tablefootmark{a} & $v \sin i$ (km/s)  & $\geq$10&  17              & 108       & 200               \\\hline
 Obtained  & $\Delta \nu$($\mu$Hz)   &  22.2 $\pm$ 0.4      & 55.4 $\pm$ 0.8    & 40.3 $\pm$ 0.6               & 55.0 $\pm$ 0.6    \\
 parameters & $s$($\mu$Hz)            &  1.2 $\pm$ 0.1       & 7.1 $\pm$ 0.2     & 8.1 $\pm$ 0.5                & 16.9 $\pm$ 0.7    \\\hline
          & $\rho(\rho_{\odot})$\tablefootmark{b}    &  0.027 $\pm$ 0.001   & 0.169 $\pm$ 0.005 & 0.089 $\pm$ 0.005            & 0.17 $\pm$ 0.01   \\
          & $M(M_{\odot})$\tablefootmark{b}          &  2.2 $\pm$ 0.3       & 1.5 $\pm$ 0.3     & 2.0 $\pm$ 0.1                & 2.4 $\pm$ 0.7     \\
          & $R(R_{\odot})$\tablefootmark{b}          &  4.2 $\pm$ 0.5       & 2.09 $\pm$ 0.07   & 2.8 $\pm$ 0.1                & 2.4 $\pm$ 0.2     \\
          & $\epsilon^{2}$(\%)\tablefootmark{c}      &  0.5 $\pm$ 0.1       & 3.0 $\pm$ 0.3     & 7.4 $\pm$ 0.9                & 17 $\pm$ 1        \\ 
          & $O$(\%)\tablefootmark{c}                 &  0.27 $\pm$ 0.03     & 1.5 $\pm$ 0.1     & 3.6 $\pm$ 0.4                & 8.0 $\pm$ 0.6     \\
 Calculated & $R_{p}(R_{\odot})$\tablefootmark{c}      &  4.2 $\pm$ 0.5       & 2.07 $\pm$ 0.07   & 2.7 $\pm$ 0.1                & 2.3 $\pm$ 0.2     \\
 parameters & $R_{e}(R_{\odot})$\tablefootmark{c}      &  4.2 $\pm$ 0.5       & 2.10 $\pm$ 0.07   & 2.9 $\pm$ 0.1                & 2.5 $\pm$ 0.2     \\
          & $i(^{o})$               &        -             & 15 - 35           & 88 - 90                      & 55 - 90           \\
          & $\Omega/\Omega_{k}$(\%) &  14 $\pm$ 7          & 31 $\pm$ 3        & 48 $\pm$ 6                   & 70 $\pm$ 7        \\ 
          & $\mathrm{log}~\Gamma$\tablefootmark{d} &  -8.14 $\pm$ 0.08   & -8.15 $\pm$ 0.04  & -8.09 $\pm$ 0.01  & -8.11 $\pm$ 0.04 \\
          & $\mathrm{log}~E_{kin}$\tablefootmark{d}  &  11.511 $\pm$ 0.004  & 12.77 $\pm$ 0.01  & 9.481 $\pm$ 0.003            & 11.57 $\pm$ 0.02  \\
          & $\delta T_{eff}$(\%)\tablefootmark{c}    &  0.13 $\pm$ 0.03     & 0.76 $\pm$ 0.08   & 0.61 $\pm$ 0.08              & 4.5 $\pm$ 0.3     \\\hline
\end{tabular}
\tablefoot{\tablefoottext{a}{The parameters of KIC~5892969 are taken from \citet{Huber2014} and those of CoRoT stars are taken from the CorotSky Database \citep{Charpinet2006}.},\tablefoottext{b}{Mean density ($\rho$), mass ($M$), and radius of a star with spherical symmetry ($R$; see Section~\ref{s:regular}).},\tablefoottext{c}{Centrifugal to gravity force ratio ($\epsilon^2$), oblateness ($O$), polar and equatorial radii of the star ($R_p$ and $R_e$), and gravity-darkening effect ($\delta T_{eff}$; see Section~\ref{ss:rot}).},\tablefoottext{d}{Convective efficiency and kinetic energy of the wave are both related to de convective layer of the star (see Section~\ref{s:dScu}).}}
\end{table*}

This method uses part of the TRUFAS algorithm, originally built to detect p-mode oscillations in solar-like stars as described by \citet{Regulo2002} and later used to find planetary photometric transits \citep{Regulo2007}. It takes advantage of the properties of the spectrum of the subspectrum  $FFT \left\{ S(\nu ) \cdot H(\nu ) \right\}$ for which the spectral signature of the $n>l$ p-modes can be considered as an equally-spaced frequency set of peaks:
\begin{equation}
 S(\nu) = \sum^{k_{f}}_{k=k_{i}} \delta(\nu-k \Delta \nu) \, ,
\label{e:pmodesignal}
\end{equation}
where $k_{i}$ and $k_{f}$ are integers with $k_{i}<k_{f}$; and the window function is
\begin{equation}
H (\nu ) =  \left\{ \begin{array}{lr} 1 & \nu_{i} \leq \nu \leq \nu_{f}  \\
0 & otherwise\,  \end{array} \right.
\label{e:hat}
\end{equation}
where $\nu_{i}$ and $\nu_{f}$ are the frequency limits.\\

It is possible to find the large separation by looking for those values with a higher number of peaks in quefrency space\footnotemark[2]\footnotetext[2]{The quefrency or frequency$^{-1}$ ($q$) is the independent variable of the power spectrum of the power spectrum.} with a significant power excess at $q=k/ \Delta \nu$. This process is repeated for values close to the frequency limits of the subspectrum, only varying by a few $\mu$Hz. The number of coincidences for each possible periodicity is then counted (see bottom left panel in Fig.~\ref{f:S_C546}).\\

Not only can the large separation be found using this method, but also other periodicities such as the rotational splitting \citep{RocaCortes2001}. The major problem arises when the sought-after periodicity is not exactly uniform (Eq.~\ref{e:pmodesignal}) because our main assumption is broken. Nevertheless, the better the SNR of the observations, the higher the departure from an uniform frequency spacing that the method will be able to accept \citep{Regulo2002}.\\

We considered a frequency range down to approximately three times the highest studied periodicity to clearly detect possible regularities. To achieve a high SNR, we built an artificial light curve. The number of highest amplitude peaks ($I$) taken into account has to reach a compromise to include the spectral regularities and to avoid powering other periodicities or noise. Looking for this compromise, we tested several values of $I$. For most cases, taking into account $I=50$ peaks allowed us to find the rotational splitting, two times its value, the large separation, and half its value (see Table~\ref{t:sdnumethod} and bottom left panel of Fig.~\ref{f:S_C546}).\\

\subsection{Echelle diagram}
\label{ss:ED}

The echelle diagram takes advantage of the regularity of the p-modes (Eq.~\ref{e:pmodesignal}) and represents the power spectrum in constant slices \citep{Grec1983}. If the value of the slice is the large separation, the modes with the same degree ($l$) and azimuthal order ($m$) will appear to be aligned (see bottom right panel in Fig.~\ref{f:S_C546}). A possible deviation from this regular pattern is produced by the departure from the asymptotic regime and/or a high rotation rate.\\

We used this property to delimit the large separation found by previous methods (see Table~\ref{t:sdnumethod}) within a given frequency range. For close values of the preliminary value, the number of consecutive modes aligned within its error is counted. The limit is found when the number of consecutive modes aligned is lower than a threshold. As happens with other methods, a high number of peaks can power other periodicities. Therefore, only the highest amplitude modes were taken into account.\\

\subsection{Results}
\label{ss:results1}

Analysing the power spectra of CID~546 with all the methods already described (see Fig.~\ref{f:S_C546}), we find the value for the large separation of $\Delta \nu = 55.4 \pm 0.8$ $\mu$Hz and a splitting of $s = 7.1 \pm 0.2$ $\mu$Hz. All methods detected the signature of the large separation and/or their split peaks as $\sim$47.8 and 61.8 $\mu$Hz. In addition, the differences between split peaks, $\sim$7.1 $\mu$Hz, are compatible with those produced by the the rotational signature or their multiples. Specifically, using the TRUFAS procedure (see bottom left panel in Fig.~\ref{f:S_C546}), strong rotational signatures are detected at 6.0 and 7.8 $\mu$Hz, and also at twice (12.2 and 15.6 $\mu$Hz) and at thrice its value (23.3 $\mu$Hz). This departure from symmetric split peaks, around 0.9 $\mu$Hz, is in agreement with a moderate rotation rate.\\

\citet{Mantegazza2012} also search for regularities in the power spectrum of CID~546. They found a value for the large separation of $\Delta \nu = 46 \pm 6$ $\mu$Hz and a splitting of $s \sim 6.7$ $\mu$Hz. The value of the large separation is based on half of the highest peak of the FFT of the power spectrum (90.3 $\mu$Hz, see Fig.~17 in their publication). Although this value is compatible with the one we found, it is centred in one of the split peaks. Nevertheless, looking at their figure it is possible to observe three peaks that are consistent with the scenario described above.\\

The stellar mass is calculated by \citet{Mantegazza2012} using a grid of models. The value they find, 2.10 to 2.18 $M_{\odot}$, is higher than ours, $1.5\pm 0.3 M_{\odot}$ (see Table~\ref{t:4dScu} and the beginning of Sect.~\ref{s:regular} for more details), but their models do not reproduce the expected limits of the modes at the same time as the observed large separation. Nevertheless, both studies find that this star has a moderate rotation rate and low inclination. This is in agreement with its low projected velocity.\\

Considering the other three stars, we find that KIC~5892969 has a low rotation rate, $\Omega / \Omega_{k} \approx 0.14$. This is confirmed by the signature of the surface rotation: two high amplitude peaks in the low frequency regime of the power spectra found at 1.235 and 2.465 $\mu$Hz with amplitudes around 100 and 500 ppm, respectively. Therefore, the values of the polar and equatorial radii are similar to the radius of a star with spherical symmetry (see Table~\ref{t:4dScu}). In addition, the values of the mass and radius are equal to those found by \citet{Huber2014}, within errors.\\

In contrast, CID~8669 shows a high projected velocity, $v \sin i \sim$200 km/s, suggesting that this star could be a fast rotator with a very high rotation rate, the same as we find with our methodology (see Table~\ref{t:4dScu}). The high oblateness of this star produces a difference of temperature between the poles and the equator of around 320 K, $\sim4.5$ \%.\\

The case of CID~3619 has to be differentiated from the others. We confirm that this star might be a hybrid star because its convective efficiency ($\mathrm{log}~\Gamma$) is higher, and the kinetic energy of the waves ($\mathrm{log}~E_{kin}$) is lower, than the typical values for \dScu stars (see Table~\ref{t:4dScu}). \citet{Claret1998} estimate that stars with this temperature can present a more efficient convective zone, and that the gravity-darkening effect is less effective, $\beta \sim 0.32$ (see Eq.~\ref{e:dteff}). The variation of its temperature with latitude is then lower than that of CID~546 although its rotation rate and oblateness are higher.\\

\section{In depth study of the "\textit{grass}"}
\label{s:grass}

Using an acoustic ray model in a uniformly rotating star, \cite{Lignieres2009} study the relation between the rotation rate and the power-spectral structure. Depending on the rotation rate regime the spectrum shows several kinds of modes such as the 2- \& 6-period island modes that are restricted inside a torus region of the star, whispering gallery modes whose ray trajectories follow the outer boundary thanks to a rotation rate that has not destroyed its torus, and chaotic modes that are produced by rays that are not constrained into a torus.\\

Ligni\`eres' \& Georgeot's results (2009) show that chaotic modes are as visible as 2-period island modes and have higher amplitudes than 6-period island modes and whispering gallery modes when the rotation rate is moderate and the star is equator-on. This is caused by a lower disc-averaging cancellation of the chaotic behaviour than of the structured behaviour. The 2-period island modes have higher amplitudes than chaotic modes when the star is pole-on. On the one hand, for lower rotation rates, only 2-period island and whispering gallery modes are present because the chaotic regions are not developed enough. On the other hand, for higher rotations rates, all modes are present except for the 6-period island mode, whose torus has been destroyed.\\

Because the four stars we are studying show different rotation rate and oblateness, our sample helps us to analyse how the power-spectral and structural parameters of \dScu stars are modified by rotation. In that way, it allows us to compare our results with those predicted by \cite{Lignieres2009}.\\

The power spectrum of a \dScu star is formed by moderate amplitude peaks grouped in bunches forming a power excess, the so-called envelope, and a high number of low amplitude peaks making a flat plateau or grass (e.g. Fig.~\ref{f:S_4dScu_546}). To define this power excess, we used the amount of energy of the observed signal carried by the wave (Eq.~\ref{e:esignal}) and we assumed that all peaks that fulfil
\begin{equation}
 E_{i} \gtrsim 0.1 \%  \, 
\label{e:envelope}
\end{equation}
are part of the envelope. We then estimated different characteristic parameters such as the energy of the power excess or the number of peaks enclosed ($N_{env}$).\\

The flat plateau is a nearly-constant amplitude and mode density regime with a significant decrease at a specific frequency \citep[e.g.][]{Poretti2009,Mantegazza2012}. We called this the cut-off frequency ($\nu_{c}$) because the higher frequency modes are possibly not reflected as a result of losing their energy through the atmosphere, as predicted for p-modes in the standard theory. The amplitude decrease of the flat plateau ends when it reaches the noise level ($A_{N}$) at the frequency we called "noise frequency" ($\nu_{N}$). We supposed that all peaks that fulfil
\begin{equation}
 E_{i} \lesssim 0.01 \%  \, 
\label{e:grasscond}
\end{equation}
are part of the grass. Therefore, we can estimate its energy, and the number of peaks that constitute the grass, $N_{grass}$. To find its characteristic parameters, we can look for the variation of the density of peaks and also the variation in amplitude with frequency.\\

We tested these two methods by comparing our results for CID~546 with those obtained by \cite{Mantegazza2012} (see Sect.~\ref{ss:ndens} and~\ref{ss:grass}). Then, we discuss the results for all the stars in Sect.~\ref{ss:results2} 

\subsection{Density of peaks}
\label{ss:ndens}

\begin{figure}[!t]
\centering
\includegraphics[width=0.5\textwidth]{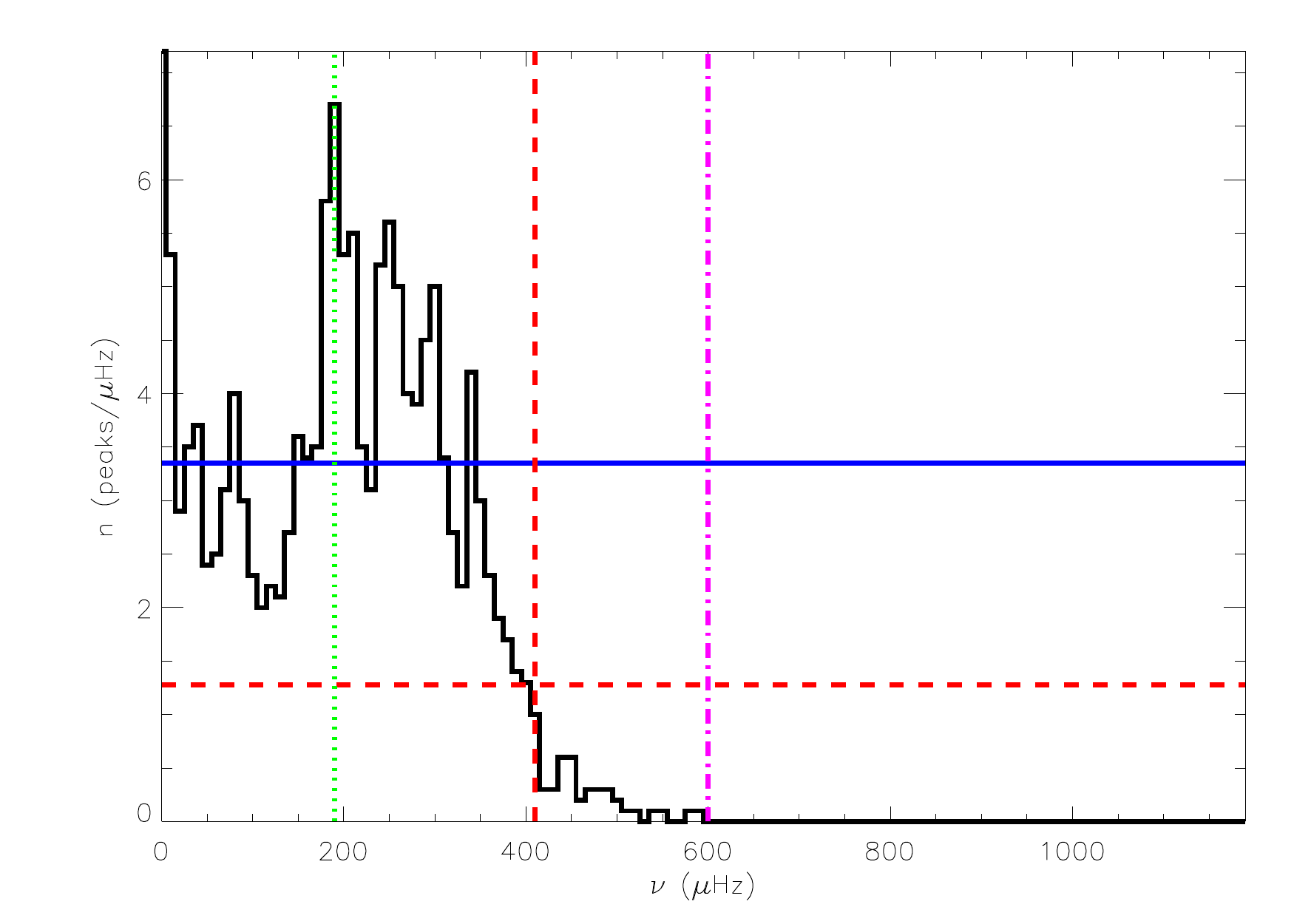}
\protect\caption[]{Density of peaks with frequency of the power spectrum of CID~546. The blue solid line indicates the mean density of peaks, the red dashed lines indicate the cut-off frequency and its density level. The green dotted and purple dashed-dotted lines indicate the frequency at maximum density and the noise frequency, respectively.}
\label{f:MoDens546}
\end{figure}

\begin{figure}[!t]
\centering
\includegraphics[width=0.5\textwidth]{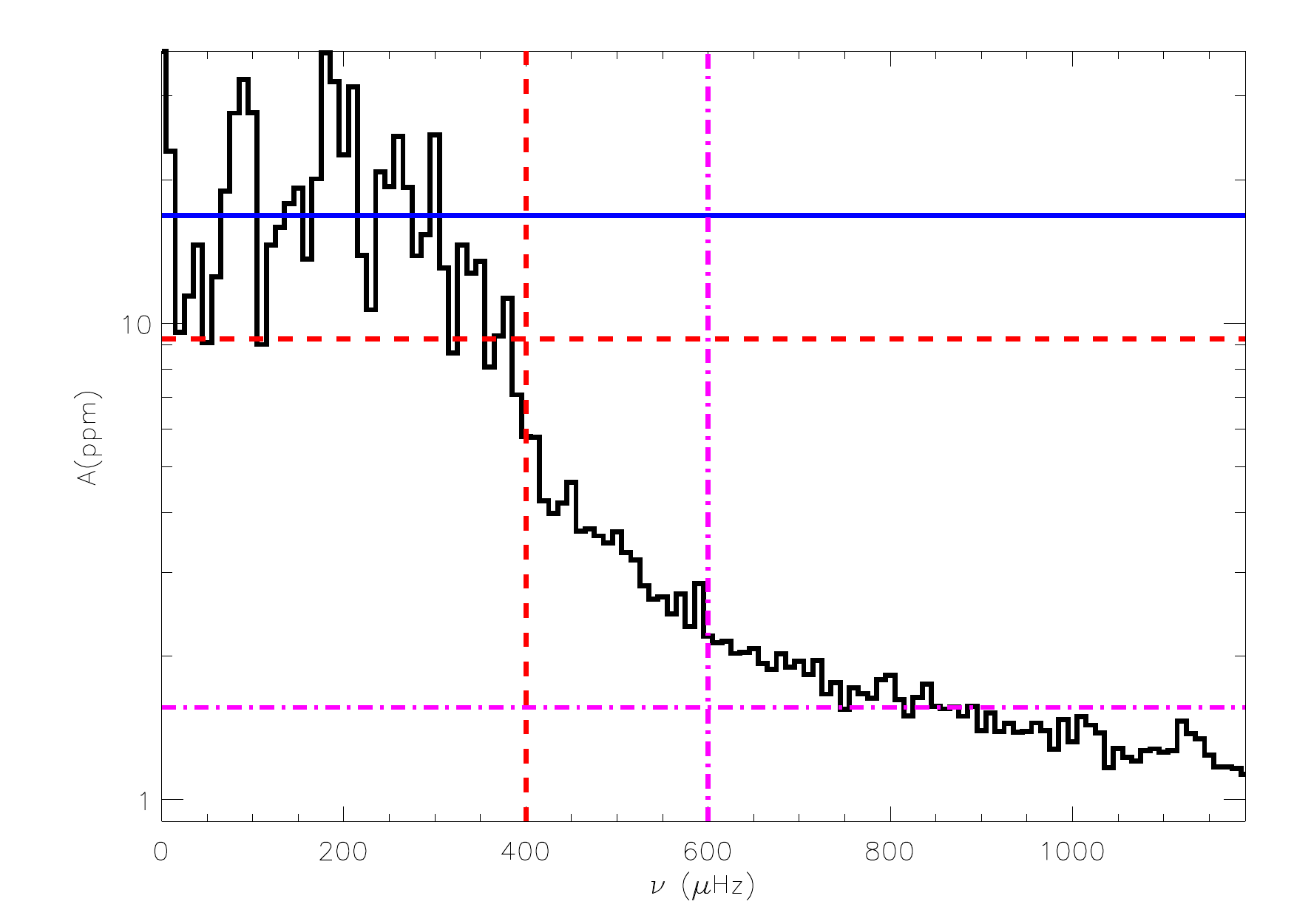}
\protect\caption[]{Mean amplitude per bin of 10 $\mu$Hz of the CID~546 power spectrum after extraction of the peaks considered as envelope. The blue solid line indicates the grass level, and the red dashed lines indicate the cut-off frequency and its amplitude level. The purple dashed-dotted lines indicate the noise frequency and the mean amplitude of noise.}
\label{f:PlatLe546}
\end{figure}

The density of peaks can be determined with a histogram of analysed peaks per 10 $\mu$Hz frequency bin (see Fig.~\ref{f:MoDens546}). The cut-off frequency can then be determined as the frequency whose density value decays more than 1.5 times the standard deviation from the mean density of peaks ($\mathbf{n}_{mean}$). We also calculated the maximum density of peaks and the frequency at maximum density. We note that the separation between higher density peaks is useful in estimating the large separation.\\

We find that the density of peaks in the power spectrum of CID~546 decays at a cut-off frequency of 405 $\pm$ 5 $\mu$Hz. This value agrees with that of \citet{Mantegazza2012} found with their analysis.\\

\subsection{Grass level}
\label{ss:grass}

Following the extraction those peaks that are considered as the envelope (see Eq.~\ref{e:envelope}) from the power spectrum, we calculated the mean amplitude of the grass or grass level ($A_{grass}$; see Fig.~\ref{f:PlatLe546}). We also find the cut-off frequency as the frequency whose amplitude value decays to more than the standard deviation from the grass level. The noise level is measured as the mean amplitude of the residual power spectrum down to the noise frequency.\\

Our analysis reveals that the grass level is an order of magnitude higher than the noise level and that the cut-off frequency is equal to 400 $\pm$ 10 $\mu$Hz. The flat plateau is clearly visible in the last two panels of Fig.~\ref{f:S_4dScu_546} after the extraction of hundreds of peaks. Our results are consistent with those found by \citet{Mantegazza2012}, because they also observe the flat plateau after the extraction of hundreds of frequencies with amplitudes down to 12 ppm.\\

\subsection{Results}
\label{ss:results2}

\begin{figure}[!t]
\centering
\includegraphics[width=0.5\textwidth]{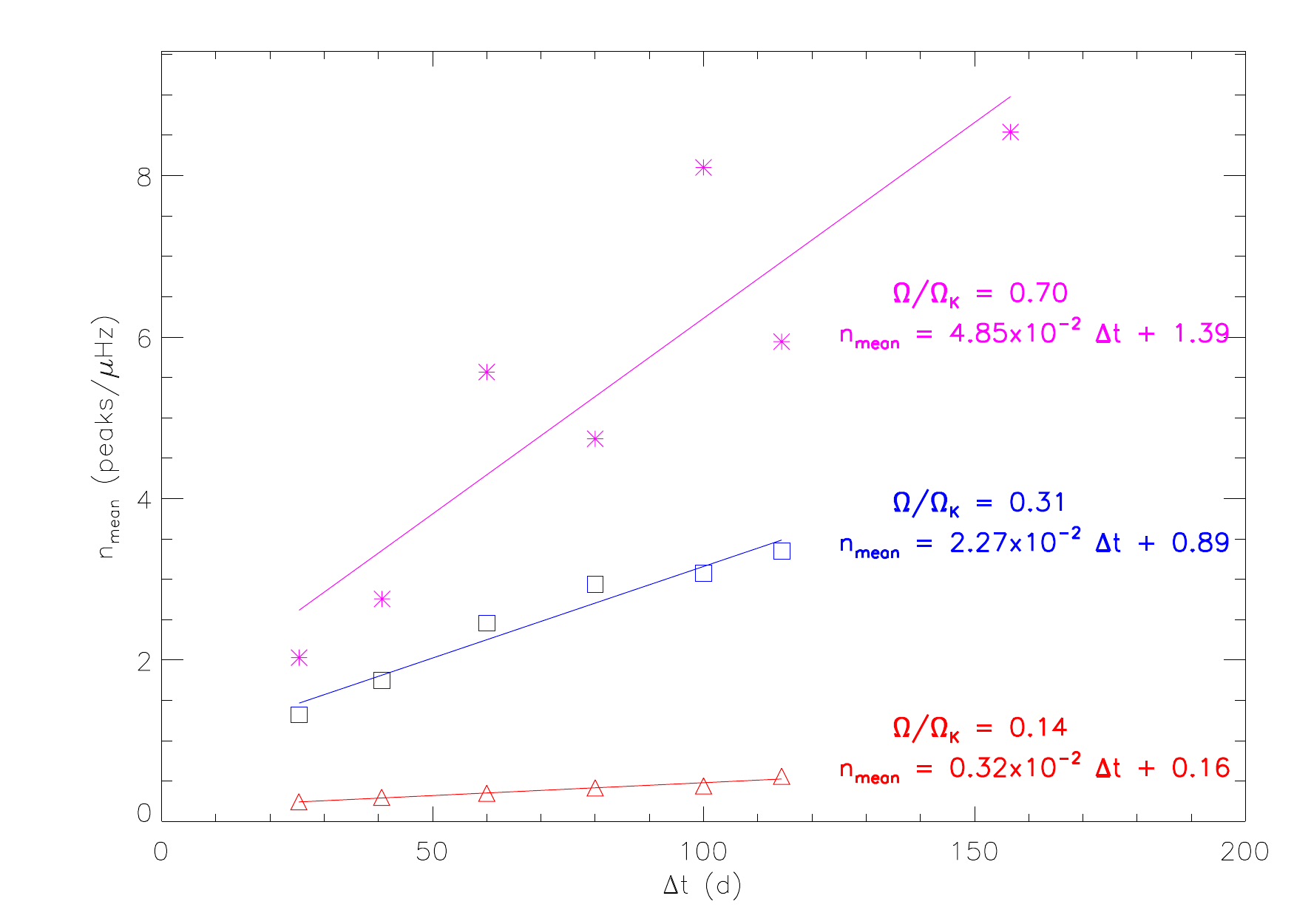}
\protect\caption[]{Mean density of peaks with duration of the studied light curves for KIC~5892969 (red triangles), CID~546 (blue squares), and CID~8669 (purple asterisks). Each line is the linear fit to the data points.}
\label{f:REvolution_Dens}
\end{figure}

\begin{figure}[!t]
\centering
\includegraphics[width=0.5\textwidth]{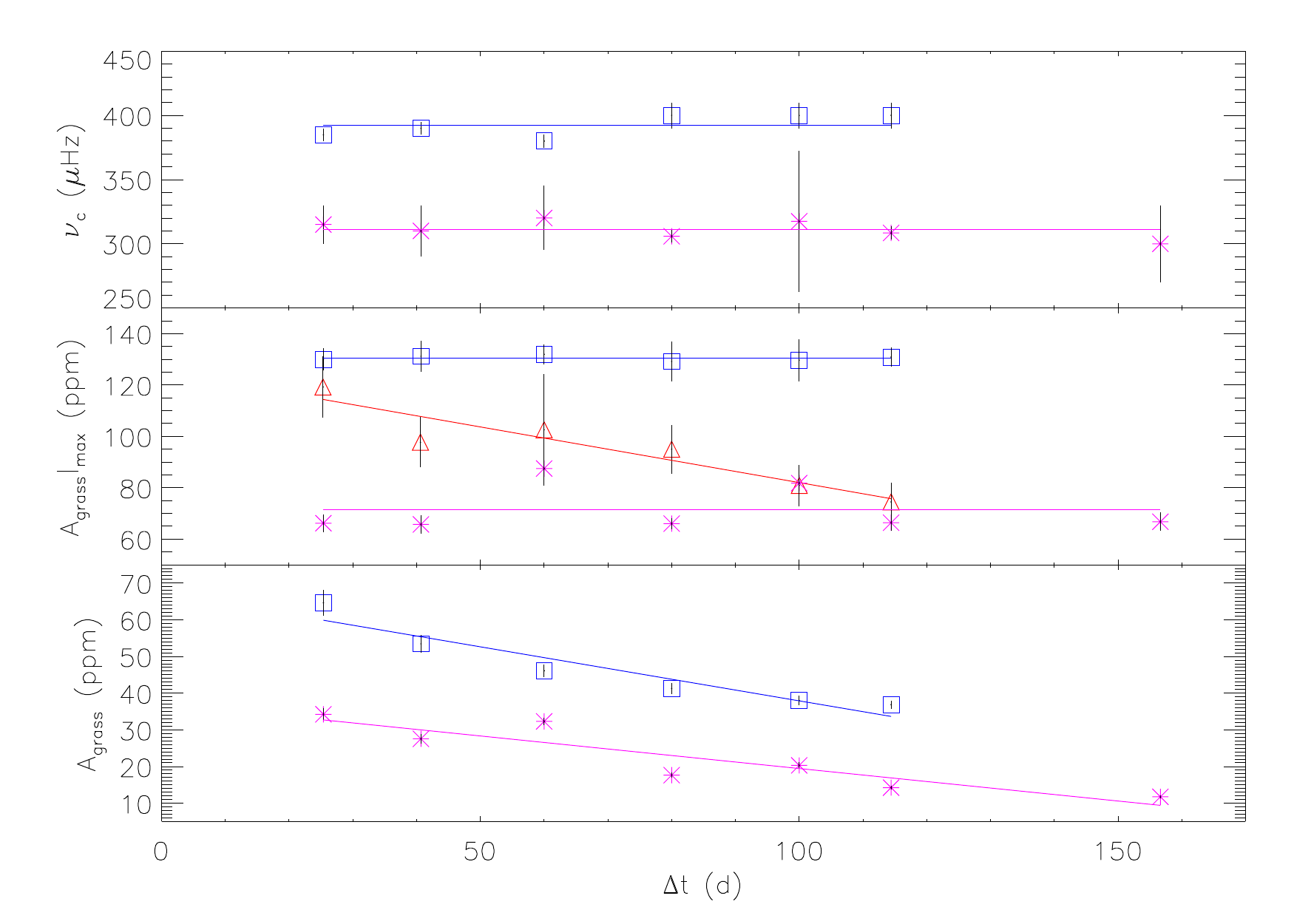}
\protect\caption[]{Cut-off frequency (top), maximum amplitude (middle), and mean amplitude of the grass (bottom panel) versus duration of the studied light curve for CID~546 (blue squares), and CID~8669 (purple asterisks). The cut-off frequency of KIC~5892969's power spectrum is not visible therefore only the maximum amplitude of the grass can be properly calculated (red triangles). Each line is the linear fit to the observed data points. The mean amplitude of the grass for CID~546 has been increased by 20 ppm to properly observe its behaviour.}
\label{f:REvolution_Plat}
\end{figure}

\begin{table*}[!t]
\caption{Comparison of the power-spectral structure of the \dScu stars studied, showing the increase of the mean density of peaks with higher rotation rate.}
\label{t:revolution}
\centering
\begin{tabular}{c c c c c c}
\hline\hline
                                  &          &Model\tablefootmark{a} &KIC~5892969 &CID~546 &CID~8669 \\
\hline
$\Omega / \Omega_{k}$& (\%)     &  59          & 14 $\pm$ 7          & 31 $\pm$ 3     & 70 $\pm$ 7      \\
$ \nu_{c}$ & ($\mu$Hz)&   & -\tablefootmark{b} & 400 $\pm$ 10     & 310 $\pm$ 7      \\
$\dot{\mathbf{n}}_{mean}$       & ($\frac{peaks\, 10^{-2}}{\mu Hz \,d}$) & & 0.32 $\pm$ 0.03 & 2.27 $\pm$ 0.25& 5 $\pm$ 1\\
$\mathbf{n}_{mean} \{0\}$& (peaks/$\mu$Hz) &        & 0.16 $\pm$ 0.03     & 0.89 $\pm$ 0.19& 1.38 $\pm$ 1.00 \\
$N_{env}$            & (peaks) & 34      & 24 $\pm$ 1\tablefootmark{c} & 17 $\pm$ 6     & 29 $\pm$ 3      \\
$N_{grass} \{0\}$    & (peaks) & 270 $\pm$ 8  & 13 $\pm$ 6\tablefootmark{b} & 282 $\pm$ 64 & 320 $\pm$ 250      \\  
\hline
\end{tabular}
\tablefoot{\tablefoottext{a}{Expected modes in the characteristic frequency range for a computed model of a star with moderate rotation rate \citep{Lignieres2009}.},\tablefoottext{b}{Since the cut off frequency is not visible, the high frequency limit to calculate the initial number of modes $N_{grass} \{0\}$ is the Nyquist frequency (see text).},\tablefoottext{c}{This mean value of $N_{env}$ does not take into account the measurement at $\Delta t=1470$ d. This duration of the light curve is long enough to observe sidelobes caused by RMC \citep[see][]{BarceloForteza2015}.}}
\end{table*}

The observed power-spectral structure of these \dScu stars consists of a few dominant amplitude modes and a lot of low amplitude peaks with the exception of CID~3619, which is a hybrid star (see Fig~\ref{f:S_4dScu_3619}). Therefore, only considering these three non-hybrid \dScu stars, CID~546, CID~8669, and KIC~5892969 together, we find that the mean density of peaks present in their power spectra increases linearly with the duration of the observing campaign ($\Delta t$; see Fig.~\ref{f:REvolution_Dens}). The density of peaks and their increase with time ($\dot{\mathbf{n}}_{mean}$) are higher as the rotation rate is higher too (see Table~\ref{t:revolution}). Therefore, the mechanism that produces this high number of peaks is related to the rotation rate, and of the increase in frequency content explains the light curve behaviour with time. Taking into account this relation, and also that the subtracted energy of the signal remains constant or slightly decreases with duration, it is not possible that all these peaks are spurious owing to an imperfect subtraction of the signals, as suggested by \cite{Balona2014a}.\\

The number of modes not caused by time variations, $N_{grass}\{0\}$, can be estimated with the y-intercept constant of the $n_{mean}$-$\Delta t$ relation while taking into account the observed frequency limits $\nu \in \left[60, \nu_{c} \right]$ $\mu$Hz (see Fig.~\ref{f:REvolution_Dens} and Table~\ref{t:revolution}). Their values are of the same order of magnitude as those chaotic modes estimated by \citet{Lignieres2009}, which take into account only axisymmetric modes in a characteristic frequency range for a computed model of a star with rotation rate around $\Omega /\Omega_{K} \sim$0.59. In addition, the observed number of modes in the envelope, $N_{env}$ (those that fulfil Eq.~\ref{e:envelope}), are also similar to those expected for 2-period island modes. As we can see, a star with higher rotation shows a higher number of chaotic modes because the torus of less-visible modes has been destroyed. The chaotic modes seem to be more visible than the 6-period island modes or the whispering gallery modes due to their irregularity, which makes the cancellation effect less effective.\\

Moreover, the maximum amplitude and cut-off frequency in CID~546 and CID~8669 are constant with time (see Fig.~\ref{f:REvolution_Plat}). As the number of lower peaks increases, the mean value of the amplitude of the grass decreases. This is also in agreement with a scenario in which initial 2-period island modes and chaotic modes with time variations are present. In agreement with the predicted visibility \citep{Lignieres2009}, CID 546 presents a similar number of 2-period island modes as the other stars in the sample, but they are of higher amplitudes due to its low inclination.\\

It is not expected that a low rotation rate \dScu star has chaotic modes. This is in agreement with the initial number of modes that we estimate for KIC~5892969. This star shows a slight decrease of its maximum amplitude of the grass. Therefore, the high number of peaks present in its power spectrum of the whole light curve can be produced by time variations of the 2-period island modes and some whispering gallery modes.\\

Finally, although CID~3619 has a higher rotation rate than CID~546, its spectral density is lower and there is not a clear flat plateau. The cause could be CID~3619's more efficient convective zone. Although \citet{Balona2014} claim that all \dScu stars are hybrids, to identify the star as a hybrid or a non-hybrid star with the criteria specified in Sect.~\ref{s:dScu} could be of importance to explaining the presence, or absence, of the flat plateau.\\

\section{Conclusions}
\label{s:conclusions}

Using our own methodology ($\delta$SBF), we analysed the light curves of four \dScu stars, observed by CoRoT and Kepler, from raw data to end products such as the parameters of the modes, the properties of the flat plateau, and possible regularities of the power spectra. We thus determine their observational characteristics producing the best estimates to date of their stellar parameters such as mass, inclination, rotation rate, and convective efficiency. In spite of the high uncertainties in previously known data, the oblateness and the gravity-darkening effect were obtained for all the stars studied. Furthermore, CID~3619 was found to be a hybrid $\delta$~Scu/$\gamma$~Dor star.\\

Because these four stars show different rotation rates and oblateness values, our sample allows us to study how the power-spectral and structural parameters of \dScu stars are modified by rotation. We prove that structural parameters such as oblateness, inclination, and convective efficiency can explain the development of the flat plateau. Therefore the power-spectral structure is formed by an envelope constituted of 2-period island modes, and a grass composed of chaotic modes and peaks due to their variation. In this sense, the spurious signal hypothesis is discarded. Our next step is to perform a study of a much larger sample of \dScu stars to provide an in-depth determination of the behaviour of their power-spectral structure.\\

\paragraph{}
\begin{acknowledgements}
The authors wish to thank the \textit{CoRoT} and \textit{Kepler} Teams whose efforts made these results possible. The \textit{CoRoT} space mission has been developed and was operated by \textit{CNES}, with contributions from Austria, Belgium, Brazil, ESA (RSSD and Science Program), Germany, and Spain. Funding for \textit{Kepler}'s Discovery mission is provided by NASA’s Science Mission Directorate. S.B.F. wishes to thank E. Michel for encouraging him to study \dScu stars, and the Solar Physics Team of the \textit{Universitat de les Illes Balears} (UIB) for hosting his stay in Majorca. He has received financial support from the Spanish Ministry of Science and Innovation (MICINN) under the grant AYA2010-20982-C02-02. A.G.H. acknowledges support from Fundação para a Ciência e a Tecnologia (FCT, Portugal) through the fellowship SFRH/BPD/80619/2011. R.A.G. acknowledges the financial support from the ANR (Agence Nationale de la Recherche, France) program IDEE (n ANR-12-BS05-0008) “Interaction Des Étoiles et des Exoplanètes” and from the CNES GOLF and PLATO grants at CEA.
\end{acknowledgements}

\bibliographystyle{aa}
\bibliography{tcb}

\appendix
\section{Obtained parameters of the oscillation modes of \dScu stars}
\label{ap:PoM}

\begin{table*}[!h]
\caption{Peaks of the CID~546 power spectrum with energies higher than 0.1\% of the full signal, which have been identified in the spectrum of the light curve.}
\label{t:peaks546}
\centering
\begin{tabular}{c c c c c c c}
\hline\hline
 Term &Frequency &Amplitude &Phase\tablefootmark{a} &Energy\\
 i&($\mu$Hz)&(ppm)&(rad)&(\%)\\\hline
 0&       198.630&       12181& -0.213& 86.13 \\
 1&       151.040&        2197&  1.584&  2.80 \\
 2&       188.078&        1879& -1.805&  2.05 \\
 3&       185.213&        1487&  2.963&  1.28 \\  
 4&        94.173&        1426& -2.382&  1.18 \\
 5&       158.108&        1219& -2.349&  0.86 \\
 6&        82.545&         908& -2.782&  0.48 \\
 7&       200.005&         836& -2.590&  0.41 \\
 8&       175.679&         907&  0.706&  0.48 \\
 9&       198.915&         642& -0.159&  0.24 \\
10&        88.663&         776&  2.009&  0.35 \\
11&       183.117&         523&  2.940&  0.16 \\
12&       183.265&         641& -1.309&  0.24 \\
13&       397.259&         476& -1.270&  0.13 \\
14&       174.049&         736&  2.657&  0.32 \\
15&       161.716&         509&  2.999&  0.15 \\
16&       162.741&         438& -3.077&  0.11 \\\hline
Error& 0.001  & 5 & 0.005 &       \\\hline
\end{tabular}
\tablefoot{\tablefoottext{a}{The phases are all with respect to the initial time of the run: $t_{J2000}$=3239.4915 d.}}
\end{table*}

We present the parameters of the highest-amplitude oscillation modes of each star that we obtain with our method. The results for KIC~5892969 were already published in \citet{BarceloForteza2015}. Because all of these modes accomplish the condition announced in Eq.~\ref{e:envelope}, they form part of the so-called envelope. Hundreds or thousands of peaks are identified with a SNR higher than four for each light curve (see from Fig.~\ref{f:S_4dScu_546} to Fig.~\ref{f:S_4dScu_5893969}).\\

We note that each oscillation mode of CID~3619's light curve has two different frequencies, one per studied run. Because these runs are separated by approximately four years, the differences in all the parameters might be caused by a modulation mechanism such as RMC. Although the observed frequency variation is of the same order of magnitude as the predicted one \citep[$\delta \nu / \nu \lesssim$0.1 \%, see][]{Moskalik1985}, it is not enough to ascertain which mechanism produces the variation of the envelope modes. Nevertheless, the cause of these variations is beyond of the scope of this work.\\

\begin{table*}[!h]
\caption{Peaks of the CID~3619 power spectrum with energies down to 0.1 \% of the full signal, which have been identified in the spectrum of both light curves. The left (right) column under each parameter corresponds to values found with SRa01 (SRa05) light curve.}
\label{t:peaks3619}
\centering
\begin{tabular}{c | c c | c c | c c | c c}
\hline\hline
Term\tablefootmark{a}&\multicolumn{2}{c}{Frequency} &\multicolumn{2}{c}{Amplitude} &\multicolumn{2}{c}{Phase\tablefootmark{b}} &\multicolumn{2}{c}{Energy}\\
i&\multicolumn{2}{c}{($\mu$Hz)}&\multicolumn{2}{c}{(ppm)}&\multicolumn{2}{c}{(rad)}&\multicolumn{2}{c}{(\%)}\\
\hline
 0&   185.755&  185.751&    430.9&  434.0&     2.23&  1.60&    23.71& 23.15 \\
 1&   111.340&  111.339&    477.3&  494.3&     2.59&  0.07&    29.10& 30.03 \\
 2&   119.308&  119.300&    246.0&  227.7&     5.37&  1.42&     7.72&  6.37 \\
 3&    15.625&   15.615&    224.8&  111.2&     4.34&  0.01&     6.48&  1.52 \\
 4&    16.101&   16.038&    231.4&  291.2&     2.84&  4.40&     6.82& 10.44 \\ 
 5&   145.069&  145.071&    171.7&  177.5&     2.03&  1.88&     3.77&  3.87 \\
 6&   118.165&  118.168&    156.1&  156.2&     0.29&  0.04&     3.11&  3.00 \\
 7&   247.298&  247.291&    133.2&  137.9&     4.77&  5.87&     2.27&  2.33 \\
 9&     9.555&    9.601&     89.0&   47.1&     0.22&  2.96&     1.01&  0.27 \\
10&    21.040&   21.114&     83.6&   55.0&     3.19&  5.59&     0.89&  0.37 \\
14&    11.266&   11.283&     74.6&  105.1&     2.15&  4.79&     0.71&  1.36 \\
15&   160.200&  160.185&     63.2&   49.1&     2.37&  0.45&     0.51&  0.30 \\
16&    14.888&   14.915&     66.9&   51.1&     4.39&  3.71&     0.57&  0.32 \\
19&    47.455&   47.331&     48.8&   33.1&     4.03&  2.91&     0.31&  0.13 \\
24&    95.656&   95.448&     38.3&   66.3&     5.44&  5.07&     0.19&  0.54 \\
25&    28.601&   28.620&     49.8&   38.8&     4.46&  2.17&     0.32&  0.19 \\
28&    22.349&   22.332&     37.1&   33.1&     2.90&  5.83&     0.18&  0.13 \\
29&    26.278&   26.405&     54.9&   56.3&     2.12&  1.51&     0.39&  0.39 \\
32&    21.598&   21.658&     32.9&   29.5&     1.29&  2.13&     0.14&  0.11 \\
33&    35.508&   35.422&     36.6&   38.7&     2.07&  5.77&     0.17&  0.18 \\\hline
Error & 0.005& 0.003 & 1.1 & 0.9 & 0.01 & 0.01 &       \\\hline
\end{tabular}
\tablefoot{\tablefoottext{a}{The terms of the modes are those used for the SRa01 light curve. The numbering is different for SRa05 since the modes have amplitude variations between each run and our method analyses the highest amplitude mode in each iteration.},\tablefoottext{b}{The phases are all with respect to the initial time of SRa01: $t_{J2000}$=2986.4802 d.}}
\end{table*}

\begin{table*}
\caption{Peaks of the CID~8669 power spectrum with energies higher than 0.1\% of the full signal.}
\label{t:peaks8669}
\centering
\begin{tabular}{c c c c c c c}
\hline\hline
Term&Frequency &Amplitude &Phase\tablefootmark{a} &Energy\\
i&($\mu$Hz)&(ppm)&(rad)&(\%)\\
\hline
 0&      176.4481&       3226&       1.249&   23.12\\
 1&      154.5399&       2952&       2.206&   19.35\\
 2&      103.6354&       2549&       0.018&   14.43\\
 3&      122.0414&       1454&       1.285&    4.69\\
 4&      110.3414&       1432&      -1.988&    4.56\\
 5&      121.1245&       1301&      -1.026&    3.76\\
 6&       89.2400&       1087&       1.117&    2.63\\
 7&      100.6208&        968&      -2.142&    2.08\\
 8&       88.6171&       1100&      -1.710&    2.69\\
 9&      145.1459&       1227&      -0.043&    3.35\\
10&      180.4720&       1136&       0.984&    2.87\\
11&      209.0796&        881&       1.313&    1.73\\
12&      100.0815&       1139&      -1.952&    2.88\\
13&      169.8589&        675&       1.512&    1.01\\
14&      120.0376&        702&       0.857&    1.10\\
15&       98.9628&        667&      -2.154&    0.99\\
16&       87.9891&        457&      -0.136&    0.46\\
17&      211.9782&        420&       0.685&    0.39\\
18&      153.6621&        413&       1.736&    0.38\\
19&       91.2896&        502&       0.614&    0.56\\
20&      117.2402&        374&      -1.658&    0.31\\
21&      100.1849&        313&       2.748&    0.22\\
23&      130.1431&        361&       2.271&    0.29\\
24&      151.2113&        402&       0.676&    0.36\\
25&      125.2902&        305&      -2.142&    0.21\\
26&      170.5608&        344&       2.131&    0.26\\
27&      104.6492&        312&       1.094&    0.22\\
28&      171.9381&        252&       1.744&    0.14\\
30&      142.2407&        232&       2.521&    0.12\\
31&      209.0220&        226&      -0.989&    0.11\\
32&      162.2456&        223&      -2.442&    0.11\\\hline
Error& 0.0007 & 3 &0.005 &       \\\hline
\end{tabular}
\tablefoot{\tablefoottext{a}{The phases are all with respect to the initial time of the run: $t_{J2000}$=2687.0916 d.}}
\end{table*}

\end{document}